\documentclass[superscriptaddress,secnumarabic,
amssymb,amsmath,nobibnotes,aps,prd,showkeys,showpacs,nofootinbib]{revtex4}%
\usepackage{graphicx}
\usepackage{epsf}
\usepackage{bm}
\usepackage{amsmath}
\usepackage{amsfonts}
\usepackage{amssymb}
\usepackage{epstopdf}
\usepackage{subfigure}
\usepackage{natbib}
\usepackage{color}%
\usepackage{hyperref}
\hypersetup{dvips,dvipdfm,linktoc=page,colorlinks=true,linkcolor=blue,citecolor=red,filecolor=magenta,urlcolor=magenta,bookmarks=true}
\providecommand{\U}[1]{\protect\rule{.1in}{.1in}}
\newcommand{\be}{\begin{equation}}
	\newcommand{\ee}{\end{equation}}

\newcommand{\mincir}{\raise
	-3.truept\hbox{\rlap{\hbox{$\sim$}}\raise4.truept\hbox{$<$}\ }}
\newcommand{\magcir}{\raise
	-3.truept\hbox{\rlap{\hbox{$\sim$}}\raise4.truept\hbox{$>$}\ }}

\begin{document}
	\title{ Dynamical stability of an interacting quintessence with varying-mass dark matter particles in Lyra manifold}
	\author{Goutam Mandal}
	\email{gmandal243@gmail.com}
	\affiliation{Department of Mathematics, University of North Bengal, Raja Rammohunpur, Darjeeling-734013, West Bengal, India.}
	\author{Sujay Kr. Biswas}
	\email{sujaymathju@gmail.com; sujay.math@nbu.ac.in}
	\affiliation{Department of Mathematics, University of North Bengal, Raja Rammohunpur, Darjeeling-734013, West Bengal, India.}
\begin{abstract}
In the background dynamics of a spatially flat FLRW model of the universe, we investigate an interacting dark energy model in the context of Lyra's geometry. Pressure-less dust is considered as dark matter, mass of which varies with time via scalar field in the sense that decaying of dark matter particles reproduces the scalar field. Here, quintessence scalar field is adopted as dark energy candidate which evolves in exponential potential. Mass of the dark matter particles is also considered to be evolved in exponential function of the scalar field. Cosmological evolution equations are studied in the framework of dynamical systems analysis. Dimension-less variables are chosen properly so that the cosmological evolution equations are converted into an autonomous system of ordinary differential equations. Linear stability is performed to find the nature of critical points by perturbing the system around the critical points in the phase space. Classical stability is also executed by finding out the speed of sound. Dynamical systems explore several viable results which are physically interested in some parameter regions. Late-time scalar field dominated attractors are found by critical points, corresponding to the accelerating universe. Scalar field-displacement vector field scaling solutions are realized that represent late time decelerated universe. Dark energy -dark matter scaling solutions are also exhibited by critical points which correspond to accelerated attractors possessing similar order of energy densities of dark energy and dark matter, that provides the possible solutions of coincidence problem.  
\end{abstract}

\maketitle
\section{Introduction}

In theory, the recently observed phenomena \cite{Riess:1998cb,Perlmutter:1998np,Ade:2015xua} of accelerated expansion of the universe can be realized by either introducing an exotic matter source in the energy momentum tensor (the right-hand side) of Einstein's Field equations (EFEs) or by altering the geometric part (the left-hand side) of EFEs. The matter with a huge negative pressure is dubbed as Dark Energy (DE) which is responsible for the present acceleration of the universe. Dark energy contributes the maximum (almost 70\%) part of the total energy density in the universe and it is becoming the dominant source of matter distribution of the universe. As a result dark energy has gained much attention gradually in cosmological studies from early inflation to late-time acceleration, although its nature is completely unknown to us. In the context of cosmological evolutionary scenario, several DE models have been investigated. Among them, Cosmological Constant ($\varLambda$) is the simplest one and a positive $\varLambda$ can take into account the recently observed acceleration. The $\varLambda$ together with Cold Dark Matter (CDM) constitutes $\varLambda$CDM model which gives the best fit to the recent observational data. However, $\varLambda$CDM has some severe theoretical issues at the interface of cosmology and particle physics, such as `cosmological constant problem' \cite{T. Padmanabhan2003,S. Weinberg1989,V. Sahni2000} and `coincidence problem' \cite{I. Zlatev1999}. Cosmological constant may be regarded as energy density of vacuum and there is a much discripancy (in energy density of $10^{123}$ orders of magnitude) in the value of $\varLambda$ acquired by observations and theoretically from quantum field theory. This discripancy is known as cosmological constant problem. In order to overcome such problem in cosmology, one can introduce the model of dark energy (dynamical dark energy model) having a time dependent density associated to a scalar field called as Quintessence. Thus, dynamical DE models based on scalar field have been widely studied in the literature. In this context, interested readers may look into the review \cite{Copeland}. On the other hand, the 'Coincidence problem' in cosmology refers to the fact that ``why the order of energy densities of DE and DM are similar today, though they scale differently in their cosmic evolution." In order to adress such problem, one can introduce an interaction in the dark sector allowing to exchange energy densities. Then, cosmological dark energy models with interactions have achieved a significant attention with growing amount of observational data. However, the nature and origin of DE and DM are unknown and they are assumed to be the dominant source of the universe, interaction between them cannot be neglected. Since there is no guiding priciple for assuming the interaction term and the coupling as well, one can choose the interactions from phenomenological perspectives. Furthermore, observational constraints can be verified by introducing a small positive coupling in interaction. Interacting scenario can provide a richer dynamics than the non-interacting one.
From the dynamical systems perspective, one can alleviate the coincidence problem by achieving critical points which satisfy \cite{Sujay2017}
$$\frac{\Omega_{DE}}{\Omega_{DM}} \approx O(1)~~ \mbox{and}~~ \omega_{eff}<-\frac{1}{3},$$ where the density parameters of DE ($\Omega_{DE}$) and DM ($\Omega_{DM}$) to be interpreted in terms of model parameters. Thus, in order to match ratio of energy densities with observations, it is only required to choice proper parameters without fine tuning the initial conditions. Cosmological dynamics of DE models including interactions are extensively studied with the help of dynamical system tools, see for instance in Refs.   \cite{Yuri.L.Bolotin2014,S.Kr.Biswas2015a,S.Kr.Biswas2015b,M.Khurshudyan2015,N.Tamanini2015,Xi-ming Chen2009,T.Harko2013,Wang2016,Copeland1,Fang1,Leon1,Honorez,Odintsov2018b,S. Bahamonde,V.K. Oikonomou,Aljaf} and in recent papers \cite{Biswas2021,Mandal2022}. Interested readers may find the extensive review on various DE models and interactions by Bahamonde {\it et.al.} in \cite{Bahamonde2018} and references therein.\\

Beside the aformentioned scenario, there is an equivalent approach to consider interacting varying-mass dark matter particles in cosmological study. The variable mass particle scenario assumes that the mass of CDM is to be time dependent through the scalar field $\phi$. As a result, dark matter energy density also becomes $\phi$ dependent. In this mechanism, DM particles acquire varying-mass depending on time via a scalar field which reproduces the scalar field dark energy \cite{Anderson}. This has theoretical justification as in this mechanism, the scalar field dependent varying-mass particles can arise from string or scalar tensor theories \cite{Damour}. However, the mass of DM particles as well as the scalar field potential can be assumed in various forms of functions depending on scalar field $\phi$, as for instance: exponential or power-law potential, exponential or power-law mass dependence etc. Interacting quintessence with varying-mass DM has been studied to find tracker solutions, age of the universe in context of FRW metric. Investigations of interacting quintessence with varying-mass DM have also been studied in Refs. \cite{Zhang,Comelli,Franca} by imposing exponential or power-law form of potential and mass dependence. Further, an interacting phantom scenario with varying-mass DM particles has been investigated in Ref.\cite{G.Leon-Saridakis} where coincidence problem cannot be alleviated or solved. Recently, an interacting phantom with varying-mass DM particles is investigated by using Center Manifold Theory in Ref. \cite{Soumya}. \\

Another approach in explaining the presently observed acceleration is to constitute modified gravity theory which in a particular limit tends to the usual general relativity. Modification of geometrical part (left hand side) of the EFEs can be executed through a suitable extension of Einstein-Hilbert action such as in $f$(R) gravity \cite{Felice}, in $f$(G) gravity \cite{Nojiri} etc. One can proceed to other geometrical modification by altering the Riemannian geometry. Weyl \cite{Weyl}, in 1918 proposed a modification of Riemannian geometry in order to unify gravitation and electromagnetism. But, Weyl's theory was unsatisfactory in physical point of view due to non-integrability of length of vector under parallel transport. Later in 1951, Lyra \cite{Lyra} proposed a modification of Riemannian geometry by introducing a guage function $A(x^{\mu})$ into the structureless manifold in which length transfers of vector are integrable and the connection is metric preserving as in Remannian geometry.
The physical frames in the Lyra geometry depend on both the coordinates and gauge function. The affine connection describing parallel transport depends on the metric and gauge function. As a result, the quantities like curvature tensor, torsion and their contracted forms will also become functions of metric as well as gauge function $A(x^{\mu})$.
In Lyra's geometry, Sen \cite{Sen1957} in 1957 formulated a field theory to obtain a static cosmological model where a vector displacement field $\psi_{\mu}$ arises naturally in Lyra formalism and is indebted for the redshift of the galactic spectral lines. 
According to the author, a displacement vector field $\beta$ comes out as a direct consequence of the $\psi_{\mu}$ in this context. A constant displacement vector field $\beta$ can mimic the cosmological constant and can provide the possible mechanism to describe late time acceleration without invoking cosmological constant term (in any ad-hoc manner) in field equation (see in Ref. \cite{Halford}) while the time-varying displacement field $\beta(t)$ can give some interesting scenarios in the cosmological point of view. For instance, the author, in Ref. \cite{Beesham} has studied the time dependent displacement vector field of Lyra geometry in context of vacuum and non-empty FLRW cosmology where he showed that the models solved the singularity, entropy and horizon problems. After that several authors have studied cosmological models in Lyra's geometry with constant as well as time dependent displacement field  \cite{Darabi,Kangujam P Singh,Hoavo,Shchigolev,Khurshudyan-Lyra}.
Recently, a scalar-tensor theory of gravity has also been studied in Ref. \cite{Cuzinatto} on Lyra manifold where a generalised Lyra invariant action has been constructed to produce the modified field equations for metric and gauge functions. It has also been shown that the model has a well defined Newtonian limit.
In Ref.\cite{Hoavo}, the authors have shown that the displacement vector when interacts with pressure-less DM can mimic the cosmological constant. However, in order to explain the inflation and late time acceleration scenario, modified gravity theories as well as dark energy models have received a lot of interests now. Thus, the study of scalar field DE models would be of great interest in the framework of Lyra's geometry which may be relevant to late time accelerating models \cite{Shchigolev}. An interacting quintessence DE model has been investigated in Lyra's manifold in the background of FRW universe in Ref. \cite{Khurshudyan-Lyra}.\\
  
Motivated from the above facts, we shall investigate dynamical systems analysis of an interacting quintessence scalar field model which interacts with pressure-less dark matter varying with time via a scalar field $\phi$ in the framework of Lyra's geometry.  Here, the displacement vector field $\beta$ is considered to be time dependent through scale factor as  $\beta(t)\propto a^{-3}(t)$. We also consider the exponential scalar field potential and exponential mass dependence (on $\phi$) of the DM particles. We then convert the cosmological evolution equations into a system of ordinary differential equations and linear stability theory is performed to understand the nature of critical points. Classical stability is also performed for the model considered. From dynamical analysis, we obtain some cosmological viable critical points which represent late time attractor solutions in quintessence era. Scaling attractors are also obtained in the phase space that exhibit the late phase of the universe alleviating coincidence problem successfully. Finally, we are able to find some parameter regions in which the individual critical point is stable locally as well as classically.\\

The paper is organized as follows: in section \ref{model and autonomous system}, we present the model of interacting quintessence in framework of Lyra's geometry and formulation of autonomous system. Cosmological parameters are also presented in terms of phase space variables therein. The section \ref{phase space autonomous system} comprises phase space analysis of the autonomous system with local stability as well as the classical stability of the points and section \ref{cosmological implications} shows the cosmological implications of the model and finally a short discussion has been made in the last section \ref{discussion}.


\section{ Model of Interacting quintessence with varying mass dark matter particles in Lyra's manifold and formulation to an autonomous system }\label{model and autonomous system}

In this section, first we shall discuss the interacting DE model with varying-mass dark matter particles in the context of Lyra's manifold and then discuss the formulation of autonomous system from the cosmological evolution equations by a suitable  transformation of variables.

\subsection{The model}

Lyra's geometry \cite{Lyra} is based on the modification of Riemannian geometry which closely resembles to Weyl's geometry. In this geometry, Lyra proposed a displacement vector between two neighbouring points $P(x^{\mu})$ and $Q(x^{\mu}+dx^{\mu})$ and it is defined by its components $A dx^{\mu}$, where $A\equiv A(x^{\mu})$ refers to the non-zero gauge function of the coordinates $x^{\mu}$. The coordinates $x^{\mu}$ and the gauge function $A$ constitute the reference frame $(A, ~x^{\mu})$ and  it obeys the following transformation rule
$$\bar{A}=\bar{A} (A, x^{\mu})~~ \mbox{and}~~ \bar{x}^{\mu}=\bar{x}^{\mu}(x^{\nu}),$$

where $\dfrac{\partial \bar{A}}{\partial A}\neq 0$ and det$\left(\dfrac{\partial \bar{x}}{\partial x}\right)\neq 0$.

Lyra \cite{Lyra} and Sen \cite{Sen1957} showed that in any general reference system, the affine connection is completely determined by the independent quantities: metric tensor $g_{\mu \nu}$ and the vector field quantity $\psi_\mu$. This vector field quantity $\psi^{\mu}=g^{\mu \nu} \psi_{\nu}$ is called the displacement vector field and it appears as a natural consequence of the introduction of gauge function $A(x^{\mu})$ into the structureless manifold. The affine connection $\tilde{\Gamma}^{\mu} _{\nu \sigma} $ on this manifold is defined by
\begin{equation}\label{key}
\tilde{\Gamma}^{\mu} _{\nu \sigma} =A^{-1} \Gamma^{\mu} _{\nu \sigma} +\frac{1}{2} (\delta^{\mu} _{\nu} \psi_{\sigma} +\delta^{\mu} _{\sigma} \psi_{\nu} -g_{\nu\sigma} \psi^{\mu}),
\end{equation}
where the connection $\Gamma^{\mu} _{\nu \sigma}$ is defined in terms of metric tensor $g_{\mu\nu}$ as in Riemannian geometry.  The curvature tensor, torsion tensor and their contracted forms will also become functions not only of metric but also of gauge function. The infinitesimal parallel transport of a vector field  $V^{\mu}$ is defined by 
\begin{equation}
\delta V^{\mu}=  \hat{\Gamma}^{\mu} _{\nu \sigma} V^{\nu} A dx^{\sigma}
\end{equation}
where
\begin{equation}\label{symmetric connection}
	\hat{\Gamma}^{\mu} _{\nu \sigma}=\tilde{\Gamma}^{\mu} _{\nu \sigma}-\frac{1}{2} \delta ^{\mu} _{\nu} \psi_{\sigma}
\end{equation}
The quantity $\hat{\Gamma}^{\mu} _{\nu \sigma}$ is not symmetric but $\tilde{\Gamma}^{\mu} _{\nu \sigma}$ is symmetric in the lower indices $\nu$ and $\sigma$ in Lyra manifold. The length of vector unlike in Weyl's geometry does not change under paraller transport. The metric (line element) in the Lyra geometry is given by 
\begin{equation}
	ds^{2}= g_{\mu\nu} A dx^{\mu} A dx^{\nu}
\end{equation}
which is invariant under both coordinates and gauge transformations.
The curvature tensor $\tilde{R} ^{\mu} _{\nu\rho\sigma}$ of Lyra geometry is defined by virtue of parallel transport of a vector along a closed curve: 
\begin{equation}\label{curvature tensor}
\tilde{R} ^{\mu} _{\nu\rho\sigma} =A^{-2} \left\{ \dfrac{\partial }{\partial x^{\rho}} (A \hat{\Gamma}^{\mu} _{\nu \sigma} ) - \dfrac{\partial }{\partial x^{\sigma}} (A \hat{\Gamma}^{\mu} _{\nu \rho} ) +A \hat{\Gamma}^{\mu} _{\lambda \rho} A \hat{\Gamma}^{\lambda} _{\nu \sigma} - A \hat{\Gamma}^{\mu} _{\lambda \sigma} A \hat{\Gamma}^{\lambda} _{\nu \rho} \right\}
\end{equation}
 where $\hat \Gamma$ is defined as in Eqn. (\ref{symmetric connection}). Cosequently, by contracting the curvature tensor in Eqn. (\ref{curvature tensor}), one obtains the curvature scalar as
 
 \begin{equation}\label{curvature in Lyra1}
 	\tilde{R}= A^{-2} R  +3 A^{-1} \psi^{\mu} _{; \mu} +\frac{3}{2} \psi^{\mu} \psi_{\mu} +2 A^{-1} \dfrac{\partial }{\partial x^{\mu}} \left\{log (A)^{2} \right\} \psi^{\mu}
 \end{equation}
where $R$ is the Riemann curvature scalar which directly depends on gauge function and semicolon stands here for covariant derivative with respect to the Christoffel symbols of the second kind in the Riemannian sense. The invariant volume integral in four dimensional Lyra manifold is given by
\begin{equation}\label{action1}
	I= \int L \sqrt{-g} A^{4} d^{4}x
\end{equation}
where $L$ is scalar and invariant in this geometry. 
Now by using the normal gauge, {\it i.e.}, $A=1$ \cite{Sen1971}  (putting in Eqn. (\ref{curvature in Lyra1})), the curvature scalar reduces to the following form:  
\begin{equation}\label{scalar curvature 2}
	\tilde{R}=R  +3  \psi^{\mu} _{; \mu} +\frac{3}{2} \psi^{\mu} \psi_{\mu}
\end{equation}
and by substituting $L=\tilde{R}$ \cite{Halford,Sen1971} in Eqn. (\ref{action1}), the action in Lyra manifold takes the form 
\begin{equation}\label{action2}
	I=\int \tilde{R} \sqrt{-g} d^{4}x
\end{equation}
Note that the action (\ref{action2}) describes the Lyra invariant action, {\it i.e.}, the Einstein Field Equations obtained through this action are invariant under both the scale and coordinate transformations. Hence, the action is a simple generalisation of Einstein-Hilbert action. 
Now the field equation may be derived from variational principle
\begin{equation}\label{Variational Principle}
	\delta (I+ I_m)=0,
\end{equation}
where $I_m$ is the action for the matter Lagrangian and is given by 
\begin{equation}\label{action3}
	I_m = \int L_{m} \sqrt{-g} d^{4} x.
\end{equation}
Thus after varying the action with respect to metric and gauge function, one obains the Einstein Field Equation in Lyra geometry.
The action (\ref{action2}) together with the action (\ref{action3}) gives the field equation (\ref{modified EFE}) via the variational principle (\ref{Variational Principle}). Here, we take the basic ingredients such as the Lyra invariant integral ({\it i.e.}, 4-dimensional volume element of space time $\sqrt{-g} A^{4} d^{4}x$) and the scalar curvature in Eqn.(\ref{scalar curvature 2}) which plays an important role as Lagrangian density.
Recently, Lyra scalar-tensor theory is investigated in Ref. \cite{Cuzinatto} where it is shown that field equations for $g_{\mu\nu}$ and $\psi_{\mu}$ are obtained from Lyra invariant action which is a direct generalisation of Einstein-Hilbert action. The field equations here have a well defined Newtonian limit, for which it can be seen that both the metric and scale function play a role in the description of gravitational interaction.

In the context of Lyra's geometry the modified field equation reduces to the form:
\begin{equation}\label{modified EFE}
	G_{\mu\nu}+\frac{3}{2}\psi_{\mu}\psi_{\nu}-\frac{3}{4}g_{\mu \nu}\psi^{\alpha}\psi_{\alpha}=T_{\mu \nu},
\end{equation} 
where $G_{\mu\nu}=R_{\mu\nu}-\frac{1}{2}Rg_{\mu\nu}$ is the Einstein tensor and $T_{\mu\nu}$ is the energy-momentum tensor for matter field with perfect fluid satisfying 
\begin{equation}\label{Energy momentum tensor}
	T_{\mu\nu}=(\rho+p)u_{\mu}u_{\nu}-pg_{\mu\nu},
\end{equation} 
where $u_\mu=(1,0,0,0)$ is the co-moving four-velocity vector with norm $u^{\mu}u_{\mu}=1$ and $\rho(t)$ and $p(t)$ are energy density and thermodynamic pressure of the matter field characterized by the equation of state parameter $\omega=\frac{p(t)}{\rho(t)}$. It should be mentioned that the Eqn.(\ref{modified EFE}) is used widely in the literature to form a model of cosmological dynamics in the framework of Lyra geometry \cite{Halford,Bhamra,Beesham2,Reddy,Singh1,Singh2}. There are two types of displacement vector fields which have been studied. One is time independent constant displacement vector
$\psi_{\mu}=(\beta,0,0,0)$ which can mimic the cosmological constant and consequently is responsible for the present acceleration of the universe. Another is time dependent displcement vector field $\psi_{\mu}=(\beta(t),0,0,0)$ has also been studied in Ref. \cite{Beesham} to solve the singularity, entropy and horizon problems.
We shall now consider the displacement vector field \cite{Khurshudyan-Lyra,Saadat}
$\psi_{\mu}=(\frac{2}{\sqrt{3}}\beta(t),0,0,0)$ in our study,
where $\beta(t)$ represents the time-like time varying displacement vector. Although, time-like constant displacement vector has been introduced in literature, we are interested here to study the model with time varying displacement vector field. Here, the factor of $\frac{2}{\sqrt{3}}$ is taken with $\beta$ for mathematical simplicity only. It is to be noted that the field equation (\ref{modified EFE}) still possesses the four-dimensional diffeomorphisms invariance. As a result, the vector field cannot be chosen by hand so that the full space-time diffeomorphisms as symmetry of equation (\ref{modified EFE}) may be destroyed. Also, one can consider a specific model with $\beta=0$ ({\it i.e.}, when $\psi_{\mu}$ vanishes) then the modified equation (\ref{modified EFE}) reduces to the Einstein Field Equations in general relativity. Therefore, gravitational theory in Lyra manifold recovers the Einstein's equations in the limiting case of $\beta=0$.
We now proceed to apply the above constructions in cosmological study. We shall consider the spatially flat, homogeneous and isotropic FLRW metric:
 \begin{equation}\label{FLRW metric}
 	ds^2=-dt^2 +a^{2}(t) \left(dr^2 +r^2 (d\theta^2 +sin^2 \theta~ d\varphi^2)\right)
  \end{equation}  
which in the context of Lyra's geometry (with Eqn.(\ref{modified EFE}) and Eqn.(\ref{Energy momentum tensor})) yields the modified Friedmann equation and acceleration equation as
 (using natural units $\left(\kappa^{2}=8\pi G=c=1  \right)$):
\begin{equation}\label{Friedmann}
3 H^{2}-\beta^{2}(t) =\rho_{\phi} +\rho_{m}
\end{equation}
and
\begin{equation}\label{Raychaudhuri}
 - 2 \dot{H}-\beta^{2}(t)-3H^{2} = p_{\phi},
\end{equation}
where $H=\frac{\dot{a}}{a}$ is Hubble parameter defined by scale factor $a(t)$. An over-dot stands for the differentiation with respect to the cosmic time t. The parameters $\theta$ and $\varphi$ in the metric (\ref{FLRW metric}) describe the usual azimuthal and polar angles of spherical coordinates with $0\leq\theta\leq\pi$ and $0\leq\varphi<2\pi$. The energy density of the dark matter is represented by $\rho_{m}$. 
The energy density $\rho_{\phi}$ and pressure $p_{\phi}$ of quinessence scalar field are defined as: 
\begin{equation}\label{energy density scalar}
	\rho_{\phi}=\frac{1}{2}\dot{\phi}^2+V(\phi)
\end{equation}\\
and 
\begin{equation}\label{pressure scalar}
	p_{\phi}=\frac{1}{2}\dot{\phi}^2-V(\phi),
\end{equation}
where $V(\phi)$ is the self interacting potential and $\frac{1}{2}\dot{\phi}^2$ is the kinetic term of the scalar field. The equation of state parameter for scalar field reads as $\omega_{\phi}= \frac{p_{\phi}}{\rho_{\phi}}$.
Here, it is assumed that the universe is dominated by dark sector which includes pressure-less dust (as DM) and quintessence scalar field (as DE). In this work, we consider the universe is filled with dark energy, dark matter and contribution of energy density from the displacement field $\beta(t)$ (arises from Lyra's geometry). If $\rho_{Total}$ and $p_{Total}$ represent the total effective energy density and effective pressure of all matter content in the universe, one can arrange the total energy density and total pressure as:
\begin{equation}\label{rho tot}
	\rho_{Total}=\rho_{m}+\rho_{\phi}+\beta^{2}(t)
\end{equation}
and
\begin{equation}\label{p tot}
	p_{Total}=p_\phi+\beta^{2}({t})
\end{equation}
which obey the continuity equation 
\begin{equation}\label{continuity}
	\dot{\rho}_{Total}+3H(\rho_{Total}+p_{Total})=0.
\end{equation}
Then, Eqn. (\ref{continuity}) can be expanded ( using Eqn. (\ref{rho tot}) and Eqn. (\ref{p tot})) in terms of energy densities of DE, DM and $\beta$ as
\begin{equation}
	\dot{\rho}_{m}+\dot{\rho_{\phi}}+2 \beta\dot{\beta}+3 H(\rho_{m}+\rho_{\phi}+\beta^{2}+p_{\phi}+\beta^{2})=0.
\end{equation}
Above equation can be re-arranged in the following form 
\begin{equation}\label{split}
	\dot{\rho}_{i}+3 H(\rho_{i}+p_{i})+2 \beta\dot{\beta}+3 H(\beta^{2}+\beta^{2})=0,
\end{equation}
where $\rho_{i}=\rho_{\phi}+\rho_{m}$ and $p_{i}=p_\phi$. Total energy density $\rho_{Total}$ and total pressure $p_{Total}$ frame the total (effective) equation of state parameter for the model as in the following
\begin{equation}\label{equations}
\omega_{eff}=\frac{p_{Total}}{\rho_{Total}}= \frac{p_{\phi}+\beta^{2}}{\rho_{m}+\rho_{\phi}+\beta^{2}}
\end{equation}
by which we can predict whether the universe is accelerating or not.
It is assumed that the displacement vector field $\beta(t)$ does not interact with other matter source (like DE, DM etc). The equation (\ref{split}), then immediately separates the following equation
\begin{equation}\label{continuity beta}
	\dot{\beta(t)}+3H\beta(t)=0.
\end{equation}
The above equation describes the conservation equation for the displacement vector field $\beta(t)$ yielding that $\beta(t)$ evolves like a fluid with $\beta(t)\propto a^{-3}(t)$. As a result, the energy density $\rho_{\beta}$ and pressure density $p_{\beta}$ for the displacement field can be defined as $\rho_{\beta}=p_{\beta}=\beta^2$. Then, equation of state parameter for this $\beta$-fluid can be formulated as $\omega_\beta=\frac{p_{\beta}}{\rho_{\beta}}=+1$ which shows that the fluid is mimicking the stiff matter. It should be noted that in case of vacuum ({\it i.e.}, $T_{\mu\nu}=0$), after taking the Bianchi identities of (\ref{modified EFE}), a first order differential equation of $\beta(t)$ arises as in Eqn. (\ref{continuity beta}). Hence, the time dependent field $\beta(t)$ cannot be chosen arbitrarily. In absence of matter field ($T_{\mu\nu}=0$)  the constraint and acceleration equations (\ref{Friedmann}) and (\ref{Raychaudhuri}) reduce to \cite{Beesham}:
	$3H^{2}=\beta^{2}(t)$ and $- 2 \dot{H}-\beta^{2}(t)-3H^{2}=0$
by which one can show that Eqn. (\ref{continuity beta}) is satisfied. So, this first order equation for $\beta(t)$ describes the well-defined system of equations (\ref{modified EFE}) for $g_{\mu\nu}$ and $\psi_{\mu}$.

 Now, after separating Eqn. (\ref{continuity beta}), the Eqn. (\ref{split}) leads to the continuity equation of dark components in dark sector as:

\begin{equation}\label{continuity DE-DM}
	\dot{\rho_i}+3H(\rho_i+p_i)=0.
\end{equation}
Now we undertake the model of varying-mass dark matter in the dark sector. For this scenario, dark matter particles depend on time `t' through scalar field $\phi$ and its number density must obey the following conservation equation:
 \begin{equation}\label{number density}
 	\dot{n_m}+3Hn_m=0.
 \end{equation}
Since the mass of DM particles $M_m(\phi)$ depends on the scalar field $\phi$, the energy density $\rho_{m}$ of DM is also a $\phi$-dependent function given by
\begin{equation}\label{energy density number density}
	\rho_{m}(\phi)=M_m (\phi) n_m.
\end{equation} 
Using Eqn.(\ref{number density}), Eqn.(\ref{energy density number density}) gives the evolution equation for $\rho_{m}(\phi)$:
\begin{equation}\label{continuity DM}
	\dot{\rho_m}+3H\rho_m=\frac{M_{m}^{'}(\phi)}{M_{m}(\phi)}\dot{\phi}\rho_{m}
\end{equation}
which is the modified conservation equation for DM, where $'\equiv \frac{d}{d\phi}$ stands for derivative with respect to scalar field $\phi$. From the energy conservation equation (Eqn.(\ref{continuity DE-DM})), conservation equation for DE takes the following form:
\begin{equation}\label{continuity DE}
	\dot{\rho_\phi}+3H(\rho_\phi+p_\phi)=-\frac{M_{m}^{'}(\phi)}{M_{m}(\phi)}\dot{\phi}\rho_{m}
\end{equation}
It is obvious from the above fact that the study of the varying-mass DM particles is equivalent to the study of the interacting DE-DM scenario with an appropriate interaction term. In this case, the term $\frac{M_{m}^{'}(\phi)}{M_{m}(\phi)}\dot{\phi}\rho_{m} $ plays a role of interaction term in dark sector and there happens energy exchange from DM to DE if $M_{m}^{'}(\phi)\dot{\phi}<0$ while energy flow occurs in the direction of DM from DE if $M_{m}^{'}(\phi)\dot{\phi}>0$.
Equivalently the above conservation equations for DM (Eqn.(\ref{continuity DM})) and DE (Eqn.(\ref{continuity DE})) can be written in terms of effective equation of state for DM and DE as follows:
\begin{equation}\label{}
	\dot{\rho_m}+3H\rho_m(1+\omega^{(m)}_{eff})=0,
\end{equation}
  and 
\begin{equation}\label{}
	\dot{\rho_\phi}+3H\rho_{\phi}(1+\omega^{(\phi)}_{eff})=0
\end{equation}
where $\omega^{(m)}_{eff}=-\frac{M_{m}^{'}(\phi)}{M_{m}(\phi)}\frac{\dot{\phi}}{3H}$ represents the effective equation of state for DM
and  $\omega^{(\phi)}_{eff}=\omega_{\phi}+\frac{M_{m}^{'}(\phi)}{M_{m}(\phi)}\frac{\dot{\phi}}{3H}\frac{\rho_{m}}{\rho_{\phi}}$ is the effective equation of state for DE.
The evolution equation of scalar field (from Eqn. (\ref{continuity DE}), using Eqn. (\ref{energy density scalar}) and Eqn. (\ref{pressure scalar})) takes the form 
\begin{equation}\label{EvolutionEqn-scalar}
  \ddot{\phi}+V'+3H\dot{\phi}=-\frac{M_{m}^{'}(\phi)}{M_{m}(\phi)}\rho_{m}
\end{equation}  
where $'\equiv \frac{d}{d\phi}$ represents the derivative with respect to scalar field $\phi$. 

\subsection{Autonomous system and cosmological parameters}\label{autonomous system and parameters}

In this subsection, we shall construct the autonomous system of ordinary differential equations from cosmological evolution equations derived in previous subsection and write the cosmological parameters in terms of dynamical variables. Dynamical variables are chosen as dimensionless and are normalized over Hubble scale. Considering exponential potential and exponential mass dependence, we have the following:
\begin{eqnarray}\label{exponential dependence}
	M_{m}(\phi)=M_{m0}~ \mbox{exp} \{-\alpha\phi\}~~ \mbox{and}~~  V(\phi)=V_{0}~\mbox{exp}\{\gamma\phi\},
\end{eqnarray}
where $M_{m0}$ and $V_{0}$ are constants and $\alpha$ and $\gamma$ are constant real parameters.
To frame the autonomous system, we choose following dimensionless variables:
\begin{equation}\label{dynamical_variables}
    x^2=\frac{\dot{\phi}^2}{6H^2},~~~y^2=\frac{V}{3H^2},~~~\mbox{and}~~~z^2=\frac{\beta^2}{3H^2}
\end{equation}\\
which are normalized over Hubble scale. 
Now, by adopting exponential potential and exponential mass dependence (in \ref{exponential dependence}), the governing equations lead to the following autonomous system of ordinary differential equations in terms of the dynamical variables ( of (\ref{dynamical_variables})) as:
\begin{eqnarray}
\begin{split}
   \frac{dx}{dN}& =\frac{3}{2} \left[ -x \left(1-x^2+y^2-z^2\right)-\sqrt{\frac{2}{3}} \gamma  y^2+\sqrt{\frac{2}{3}} \alpha  \left(1-x^2-y^2-z^2\right)\right]  ,& \\
   \frac{dy}{dN}& =\frac{3}{2} y \left[1+x^2-y^2+z^2+\sqrt{\frac{2}{3}} \gamma  x\right] ,& \\
   \frac{dz}{dN} & =\frac{3}{2} z \left[-1+x^2-y^2+z^2\right],
    &~~\label{autonomous_system}
\end{split}
\end{eqnarray}
where $ N=\ln a $ is the e-folding parameter taken to be independent variable.
As a result, the cosmological parameters associated to this model can immediately be interpreted in terms of the dynamical variables as follows :\\
The density parameters for quintessence scalar field (DE), for dark matter and for displacement vector field $\beta$ are 
\begin{equation}\label{density_parameter}
\Omega_{\phi}=x^2+y^2,
\end{equation}
\begin{equation}\label{density_parameter_m}
\Omega_{m}=1-x^2-y^2-z^{2},
\end{equation}
and
\begin{equation}\label{density_parameter_f}
\Omega_{\beta}=z^{2}
\end{equation} 
respectively.
The effective equation of state parameter for quintessence scalar field (DE) is
\begin{equation}\label{eqn_of_state_parameter}
\omega^{(\phi)}_{eff}=\frac{x^2-y^2}{x^2+y^2}-\frac{\sqrt{\frac{2}{3}} \alpha  x \left(1-x^2-y^2-z^2\right)}{x^2+y^2}
\end{equation}
and the effective equation of state parameter for dark matter is
\begin{equation}
\omega^{(m)}_{eff}=\sqrt{\frac{2}{3}} \alpha  x.
\end{equation}
The effective equation of state parameter for the model reads as
\begin{equation}\label{eff_eqn_of_state_parameter}
\omega_{eff}=x^{2}-y^{2}+z^{2},
\end{equation}
and the deceleration parameter for the model takes the form
\begin{equation}\label{dec_parameter}
q=-1+\frac{3}{2}(1+\omega_{eff})
\end{equation}
which shows that the condition for acceleration is :~~~$q<0$~~~{\it i.e.}~~~ $\omega_{eff}<-\frac{1}{3}$
and for deceleration is:~~~$q>0$~~~{\it i.e.}~~~ $\omega_{eff}>-\frac{1}{3}.$
Friedmann equation (\ref{Friedmann}) represents the constraint equation for the model in terms of the DM density parameter
\begin{equation}\label{density_dm}
	\Omega_{m}=1-x^{2}-y^2-z^2.
\end{equation}
Now, energy condition $ 0\leq \Omega_{m} \leq 1 $ gives the constraints in physical region with dynamical variables, called phase space boundary:
\begin{equation}
	0\leq x^{2}+y^2+z^2 \leq 1
\end{equation} 


\section{Phase space analysis of autonomous system (\ref{autonomous_system}):}
\label{phase space autonomous system}
In this section, we shall extract the critical points from the $3D$ autonomous system (\ref{autonomous_system}) and we shall investigate the local  stability (from Hartman-Grobman theory) as well as the classical stability (from sound speed) of the present model in two different subsections.\\

The critical points for the system (\ref{autonomous_system}) are the following
{\bf 
\begin{itemize}
\item  I.  Critical Point : $ A=(\sqrt{\frac{2}{3}} \alpha ,0,0)$
\item  II. Set of critical Point : $ B=(\sqrt{1-z_{c}^2},0,z_{c})$

\item  III. Set of critical Point : $C=(-\sqrt{1-z_{c}^2},0,z_{c})$

\item  IV. Critical Point : $ D=\left(-\frac{\gamma }{\sqrt{6}},\sqrt{1-\frac{\gamma ^2}{6}},0 \right)$

\item  V. Critical Point : $ E=\left(-\frac{\gamma }{\sqrt{6}},-\sqrt{1-\frac{\gamma ^2}{6}},0 \right)$

\item  VI. Critical Point : $ F=\left(-\frac{\sqrt{6}}{2(\alpha +\gamma)},\frac{\sqrt{ \alpha  (\alpha +\gamma )+\frac{3}{2}}}{ (\alpha +\gamma )},0 \right)$

\item  VII. Critical Point : $ G=\left(-\frac{\sqrt{6}}{2(\alpha +\gamma)},-\frac{\sqrt{ \alpha  (\alpha +\gamma )+\frac{3}{2}}}{ (\alpha +\gamma )},0 \right)$
\end{itemize}

}
Here, $\alpha$ and $\gamma$ are free parameters of the model and $z_c$ takes any real values in the interval $[0,~ 1]$.
Critical points and their corresponding physical parameters are presented in the table (\ref{physical_parameters}).


\subsection{LOCAL STABILITY CRITERIA OF CRITICAL POINTS :}\label{local stability}
To find local stability of the critical points, we have to  perturb the system up-to first order about the critical points ({\it i.e.,} to linearize the system around the points). Then, we need to find out the eigenvalues of perturbed matrix that can determine the stability of the critical points. Stability criteria for points of the three dimensional autonomous system (\ref{autonomous_system}) are presented in this subsection. 

\begin{itemize}
	\item 

Critical point $A$ corresponds to matter-scalar field scaling solution and it exists for $-\sqrt{\frac{3}{2}}\leq \alpha \leq \sqrt{\frac{3}{2}}$ in the phase space $x-y-z$. Contribution of energy density of displacement field $\beta$ through Lyra's manifold is absent (since, the dimensionless density parameter related to displacement vector is zero, {\it i.e.}, $\Omega_{\beta}=z^2 =0$).
The ratio of energy densities of DE and DM is defined by $r=\frac{\Omega_{\phi}}{\Omega_{m}}=\frac{2\alpha^2}{3-2\alpha^2}.$ This indicates that the critical point $A$ becomes absolutely DE dominated ($\Omega_{\phi}\approx1$) solution when $\alpha \longrightarrow \pm \sqrt{\frac{3}{2}}$. On the other hand, the point is of DM dominated ($\Omega_{m}\approx1$) for $\alpha\longrightarrow0$. The restriction on decelerating parameter $q\geq \frac{1}{2}$ (see Table \ref{physical_parameters}) demands that there always exists a decelerating universe near the critical point $A$. The dark matter effective equation of state and the dark energy effective equation of state  $\omega^{(\phi)}_{eff}=\omega^{(m)}_{eff}=\frac{2\alpha^2}{3}$ reveal that they evolve similarly. As a result, both the components behave as dust for $\alpha=0$ and stiff fluid for $\alpha=\pm \sqrt{\frac{3}{2}}$. Eigenvalues of linearized Jacobian matrix at the point $A$ are: $\left\{\lambda_{1(A)}=\gamma \alpha+\frac{3}{2}+\alpha^2,~~\lambda_{2(A)}=\alpha^2 -\frac{3}{2},~~\lambda_{3(A)}=\alpha^2 -\frac{3}{2}  \right\}$.

The point is a stable attractor either for $\left(-\sqrt{\frac{3}{2}}<\alpha <0~~\mbox{and}~~ \gamma >\frac{-2 \alpha ^2-3}{2 \alpha }\right)$~~
\mbox{or for}~~ \\
$\left(0<\alpha <\sqrt{\frac{3}{2}}~~\mbox{and}~~ \gamma <\frac{-2 \alpha ^2-3}{2 \alpha }\right)$. Therefore,  the critical point $A$ corresponds to a late time decelerated matter-scalar field scaling attractor in the above parameter region (see in fig.\ref{fig:stability_A}). A cosmological viable scenario is achieved by the critical point when it comes into complete DM dominance for $\alpha=0$. For this case, the point shows decelerated dust dominated ($\Omega_{m}=1, \Omega_{\phi}=0, \omega_{eff}=0, q=\frac{1}{2}$) universe which has saddle-like transient nature (since one eigenvalue is positive) and that can be an intermediate phase (see fig. \ref{fig:stable_DE_2D},\ref{fig:stable_DE_3D}).\\

\item
Non-isolated set of critical points represented by $B$ which exists for all $ z_{c} \in [0,~1]$ and for all real parameters $\alpha$ and $\gamma$. This set is dominated by both the scalar field (kinetic part) DE and the displacement vector $\beta$-fluid due to Lyra's manifold. The ratio of dimensionless density parameters of scalar field and displacement vector $\beta$ is $\frac{\Omega_{\phi}}{\Omega_{\beta}}=\frac{1-z^2_c}{z^2_c}$. 
Thus, the set corresponds to DE dominated solution for $z_c=0$ while it represents a complete $\beta-$ fluid dominated solution for $z_c=1$.  Scalar field behaves as stiff fluid (since $\omega^{(\phi)}_{eff}=1$). Acceleration near the set is not possible (since $\omega_{eff}=1$ and $q=2$ (see in Table \ref{physical_parameters})). Eigenvalues of linearized Jacobian matrix for the set of point $B$ are: $\left\{\lambda_{1(B)}=3+\frac{\sqrt{6}\gamma}{2} \sqrt{1-z^2_c},~~\lambda_{2(B)}=3-\sqrt{6}\alpha \sqrt{1-z^2_c},~~\lambda_{3(B)}=0  \right\}$. Here, vanishing eigenvalue ($\lambda_{3(B)}=0$) indicates that the set $B$ is  non-hyperbolic in nature. Also, the non-isolated set $B$  having exactly one vanishing eigenvalue confirms that it is a normally hyperbolic set \cite{Coley,Sujay2017} and the stability of this set can be determined by finding out the sign of remaining non-vanishing eigenvalues. Thus, the condition for the set to be stable is $0\leq z_c<1,~ \alpha >\sqrt{\frac{3}{2}} \sqrt{\frac{1}{1-z^{2}_c}}~\mbox{and}~ \gamma <-\sqrt{6} \sqrt{\frac{1}{1-z^{2}_c}}$. See fig. \ref{fig:stable_B}, where $B$ is a stable attractor solution in $x-z$ plane. Note that a particular point with coordinate $(1,0,0)$ on this set represents a scalar field dominated decelerated solution (when $z_c=0$) which is source (past attractor) for $\gamma>-\sqrt{6}$ and $\alpha<\sqrt{\frac{3}{2}}$. On the other hand, a specific point with coordinate $(0,0,1)$ (for $z_c=1$) describes a  complete  Lyra's displacement vector dominated solution ($\Omega_{\beta}=1$) and it corresponds to a decelerated past attractor (source) in the phase space (see figs. \ref{fig:stable_DE_2D},\ref{fig:stable_DE_3D}, \ref{fig:stable_FG_2D} and \ref{fig:stable_FG_3D}). From the above discussions, the set of points $B$ exhibiting a scaling solution in the phase space is always decelerated in nature whether it is scalar field dominated or $\beta$ fluid dominated.\\

\item
Another set of critical point $C$ is also a non-isolated set which exists for $0\leq z_{c}\leq1$ and for all $\alpha$, $\gamma$. The set has identical nature with the set $B$ in the phase space. It is also dominated by kinetic part of scalar field DE and displacement field $\beta$ with ratio of energy densities $\frac{\Omega_{\phi}}{\Omega_{\beta}}=\frac{1-z^2_c}{z^2_c}$. There exists always deceleration near the set of points since $\omega_{eff}=1,~ q=2$. Eigenvalues of the linearized Jacobian matrix for this set are: $\left\{\lambda_{1(C)}=3+\sqrt{6}\alpha \sqrt{1-z^2_c}
,~~\lambda_{2(C)}=3-\frac{\sqrt{6}\gamma}{2} \sqrt{1-z^2_c},~~\lambda_{3(C)}=0  \right\}$. This non-hyperbolic set of critical points having exactly one vanishing eigenvalue is normally hyperbolic set and it is stable for $0\leq z_c<1,~ \alpha <-\sqrt{\frac{3}{2}} \sqrt{\frac{1}{1-z^{2}_c}}~\mbox{and}~ \gamma >\sqrt{6} \sqrt{\frac{1}{1-z^{2}_c}}$. Fig. \ref{fig:stable_C} shows the stability of $C$ in $x-z$ plane. Therefore, the set $C$ can be a stable attractor dominated by scalar field for $z_c=0$ and dominated by displacement vector of Lyra's manifold for $z_c=1$. However, it represents always decelerated expansion of the universe. Specifically, for $z_c=0$, the set becomes a point with coordinate $(-1,0,0)$ which is a past attractor dominated by scalar field ($\Omega_{\phi}=1$) for $\alpha>-\sqrt{\frac{3}{2}}$ and $\gamma<\sqrt{6}$ (because this non-hyperbolic set has 2D unstable sub-manifold associated with two non-zero positive eigenvalues). On the other hand, for $z_c=1$, the set becomes a point with coordinate $(0,0,1)$ corresponding to a decelerated solution dominated by displacement vector of Lyra's manifold.  For this case, the point is always a past attractor (source) since there exists 2D unstable sub-manifold associated with two non-zero positive eigenvalues (see figs. \ref{fig:stable_DE_2D},\ref{fig:stable_DE_3D}, \ref{fig:stable_FG_2D} and \ref{fig:stable_FG_3D}).  \\

\item
Critical points $D$ and $E$ are same in all respect and they exist for the restriction: $-\sqrt{6}\leq \gamma \leq \sqrt{6}$ in phase space. Both the points represent the scalar field (DE) dominated solutions (since $\Omega_{\phi}=1,~\Omega_{m}=0,~ \Omega_{\beta}=0$, see Table \ref{physical_parameters}). DE behaves as any perfect fluid model (since the effective equation of state for DE is $\omega^{(\phi)}_{eff}=\frac{\gamma^2}{3}-1$) and depending upon $\gamma$, it behaves as quintessence or cosmological constant like fluid or any other exotic fluid. In particular, the DE behaves as quintessence for $0<\gamma^2<2$ and cosmological constant for $\gamma=0$.  On the other hand, the DE behaves as any exotic type fluid for $2<\gamma^2<3$ but it can never behave as phantom fluid. There exists an accelerating universe near the points in the parameter interval $-\sqrt{2}<\gamma <\sqrt{2}$. Eigenvalues of linearized Jacobian matrix for both the points $D$, $E$ are: $\left\{\lambda_{1(D,E)}=\gamma \alpha+\gamma^2 -3
,~~\lambda_{2(D,E)}=\frac{\gamma^2}{2}-3,~~\lambda_{3(D,E)}=\frac{\gamma^2}{2}-3  \right\}$. The points are hyperbolic in nature since all the eigenvalues are non-zero. They can be non-hyperbolic for $\gamma=\pm \sqrt{6}$.
The hyperbolic points are stable for the following conditions:\\
(i)~$ \alpha \leq -\sqrt{\frac{3}{2}}~~\mbox{and}~~ -\frac{\sqrt{\alpha ^2+12}}{2}-\frac{\alpha }{2}<\gamma <\sqrt{6} $ ,~or\\
(ii)~$ -\sqrt{\frac{3}{2}}<\alpha \leq \sqrt{\frac{3}{2}}~~\mbox{and}~~ -\frac{\sqrt{\alpha ^2+12}}{2}-\frac{\alpha }{2}<\gamma <\frac{\sqrt{\alpha ^2+12}}{2}-\frac{\alpha }{2} $ ,~or\\
(iii)~$ \alpha >\sqrt{\frac{3}{2}}~~\mbox{and}~~ -\sqrt{6}<\gamma <\frac{\sqrt{\alpha ^2+12}}{2}-\frac{\alpha }{2}$. \\
See figs. \ref{fig:stable_DE_2D}, \ref{fig:stable_DE_3D}, where $D$ and $E$ are stable attractor solutions in late times.
Now the points represent the late time attractors in quintessence era (with $-1<\omega_{eff}<-\frac{1}{3}$) for the following restrictions:\\
(i)~$\alpha\leq-\frac{1}{\sqrt{2}}~~ \mbox{and}~~ (-\frac{\alpha}{2}-\frac{\sqrt{\alpha^2 +12}}{2}<\gamma<0 ~~\mbox{or}~~ 0<\gamma<\sqrt{2})$ ,~or\\
(ii)~$-\frac{1}{\sqrt{2}}<\alpha\leq \frac{1}{\sqrt{2}}~~\mbox{and}~~ (-\sqrt{2}<\gamma<0~~\mbox{or}~~0<\gamma<\sqrt{2})$ ,~or\\
(iii)~$\alpha>\frac{1}{\sqrt{2}}~~\mbox{and}~~(-\sqrt{2}<\gamma<0~~\mbox{or}~~0<\gamma<-\frac{\alpha}{2}+\frac{\sqrt{\alpha^2 +12}}{2})$.

Also, the points are stable solutions attracted by cosmological constant (with $\omega_{eff}=q=-1$) for $\gamma=0$. That is for constant potential of scalar field ($\gamma=0$), the points can describe the late time accelerated de Sitter attractor solutions with $\Omega_{\phi}=1,~ \Omega_{m}=0,~ \Omega_{\beta}=0,~ \omega_{eff}=-1,~ q=-1$ and for this case scalar field DE behaves as cosmological constant like fluid $\omega^{(\phi)}_{eff}=-1$. So, one can conclude that the scalar field dominated solutions, namely, critical points $D$ and $E$ are physically interesting in late times as they correspond to accelerated stable attractors in quintessence era for a non-constant potential ($\gamma\neq 0$) and the accelerated de Sitter universe for constant potential.  

\item
Finally, the existence for the dark matter-scalar field scaling solutions represented by critical points $F$ and $G$ are following:\\
(i)~$ \alpha <-\sqrt{\frac{3}{2}}~~\mbox{and}~~ \gamma \leq -\frac{\sqrt{\alpha ^2+12}}{2}-\frac{\alpha }{2} $ ,~or\\
(ii)~$ \alpha =-\sqrt{\frac{3}{2}}~~\mbox{and}~~ \left(\gamma \leq -\sqrt{\frac{3}{2}}~~\mbox{or}~~ \gamma =\sqrt{6}\right) $ ,~or\\
(iii)~$ -\sqrt{\frac{3}{2}}<\alpha <0~~\mbox{and}~~ \left(\gamma \leq -\frac{\sqrt{\alpha ^2+12}}{2}-\frac{\alpha }{2}~~\mbox{or}~~ \frac{\sqrt{\alpha ^2+12}}{2}-\frac{\alpha }{2}\leq \gamma \leq \frac{-2 \alpha ^2-3}{2 \alpha }\right) $ ,~or\\
(iv)~$ \alpha =0~~\mbox{and}~~ \left(\gamma \leq -\sqrt{3}~~\mbox{or}~~ \gamma \geq \sqrt{3}\right) $ ,~or\\
(v)~$ 0<\alpha <\sqrt{\frac{3}{2}}~~\mbox{and}~~ \left(\frac{-2 \alpha ^2-3}{2 \alpha }\leq \gamma \leq -\frac{\sqrt{\alpha ^2+12}}{2}-\frac{\alpha }{2}~~\mbox{or}~~ \gamma \geq \frac{\sqrt{\alpha ^2+12}}{2}-\frac{\alpha }{2}\right) $ ,~or\\
(vi)~$ \alpha =\sqrt{\frac{3}{2}}~~\mbox{and}~~ \left(\gamma =-\sqrt{6}~~\mbox{or}~~ \gamma \geq \sqrt{\frac{3}{2}}\right) $ ,~or\\
(vii)~$ \alpha >\sqrt{\frac{3}{2}}~~\mbox{and}~~ \gamma \geq \frac{\sqrt{\alpha ^2+12}}{2}-\frac{\alpha }{2}.$ \\
The ratio of density parameters for DE and DM is $r=\frac{\Omega_{\phi}}{\Omega_{m}}=\frac{\alpha  (\alpha +\gamma )+3}{\gamma  (\alpha +\gamma )-3}$. The DE associated to the points $D$ and $E$ behaves as any perfect fluid nature. Depending on parameters ($\alpha$ and $\gamma$) restrictions, it can behave as quintessence, cosmological constant or a phantom fluid. However, within the existence criteria of the critical points, the DE can mimic only the quintessence ($-1<\omega^{(\phi)}_{eff}<-\frac{1}{3}$) in its evolution  for some parameter restrictions. The restrictions are
(i)~$(\alpha <-\frac{1}{\sqrt{2}}~~\mbox{and}~~ 2 \alpha <\gamma \leq -\frac{\sqrt{\alpha ^2+12}}{2}-\frac{\alpha }{2})$,~~or~~
(ii)~ $(\alpha >\frac{1}{\sqrt{2}}~~\mbox{and}~~ \frac{\sqrt{\alpha ^2+12}}{2}-\frac{\alpha }{2}\leq \gamma <2 \alpha )$.\\
Note that the above restrictions lead to the condition for acceleration of the universe also (since the effective equation of state for DE and effective equation of state for the model are same for the points {\it i.e.}, $\omega^{(\phi)}_{eff}=\omega_{eff}=-\frac{\alpha}{\alpha+\gamma}$).

Eigenvalues of linearized Jacobian matrix for both the points $F$ and $G$ are:
$\lambda_{1(F,G)}=-\frac{3}{2}\frac{(\gamma+2\alpha)}{(\gamma+\alpha)}$,~
$\lambda_{2(F,G)}=\frac{-6\alpha-3\gamma+\sqrt{-48\alpha \gamma^3 -(96\alpha^2 +63)\gamma^2 -(48\alpha^3-108\alpha)\gamma+180\alpha^2 +216}}{4\gamma+4\alpha}$,~and 
$\lambda_{3(F,G)}= \frac{-6\alpha-3\gamma-\sqrt{-48\alpha \gamma^3 -(96\alpha^2 +63)\gamma^2 -(48\alpha^3-108\alpha)\gamma+180\alpha^2 +216}}{4\gamma+4\alpha} $.\\
The hyperbolic type (since all the eigenvalues are non-vanishing) critical points $F$ and $G$ are stable for:\\
(i)~$ \alpha \leq -\sqrt{\frac{3}{2}}~~\mbox{and}~~ \gamma <-\frac{\sqrt{\alpha ^2+12}}{2}-\frac{\alpha }{2} $ ,~or\\
(ii)~$-\sqrt{\frac{3}{2}}<\alpha <0~~\mbox{and}~~ \left(\gamma <-\frac{\sqrt{\alpha ^2+12}}{2}-\frac{\alpha }{2}~~\mbox{or}~~ \frac{\sqrt{\alpha ^2+12}}{2}-\frac{\alpha }{2}<\gamma <\frac{-2 \alpha ^2-3}{2 \alpha }\right)  $ ,~or\\
(iii)~$\alpha =0~~\mbox{and}~~ \left(\gamma <-\sqrt{3}~~or~~ \gamma >\sqrt{3}\right)  $ ,~or \\
(iv)~$ 0<\alpha <\sqrt{\frac{3}{2}}~~\mbox{and}~~ \left(\frac{-2 \alpha ^2-3}{2 \alpha }<\gamma <-\frac{\sqrt{\alpha ^2+12}}{2}-\frac{\alpha }{2}~~\mbox{or}~~ \gamma >\frac{\sqrt{\alpha ^2+12}}{2}-\frac{\alpha }{2}\right) $ ,~or \\
(v)~$ \alpha \geq \sqrt{\frac{3}{2}}~~\mbox{and}~~ \gamma >\frac{\sqrt{\alpha ^2+12}}{2}-\frac{\alpha }{2} $ .\\
Both the critical points represent the stable solutions in quintessence era for the following parameter restrictions:\\
(i)~$\alpha<-\frac{1}{\sqrt{2}}~~\mbox{and}~~2\alpha<\gamma<-\frac{\alpha}{2}-\frac{\sqrt{\alpha^2 +12}}{2}$ ,~or\\
(ii)~$\alpha>\frac{1}{\sqrt{2}}~~\mbox{and}~~-\frac{\alpha}{2}+\frac{\sqrt{\alpha^2 +12}}{2}<\gamma<2\alpha$.
Universe near the critical points (within existence) can never be attracted with cosmological constant or in phantom regime. Therefore, the critical points $F$ and $G$ depict the accelerated scaling attractors in quintessence era with $\omega^{(m)}_{eff}=\omega^{(\phi)}_{eff}=-\frac{\alpha}{\gamma+\alpha}$ (see figs. \ref{fig:stable_FG_2D}, \ref{fig:stable_FG_3D}).

\end{itemize}
\subsection{Model with $\beta=0$ limiting case:}
\begin{figure}
	\centering
	\subfigure[]{%
		\includegraphics[width=7cm,height=7cm]{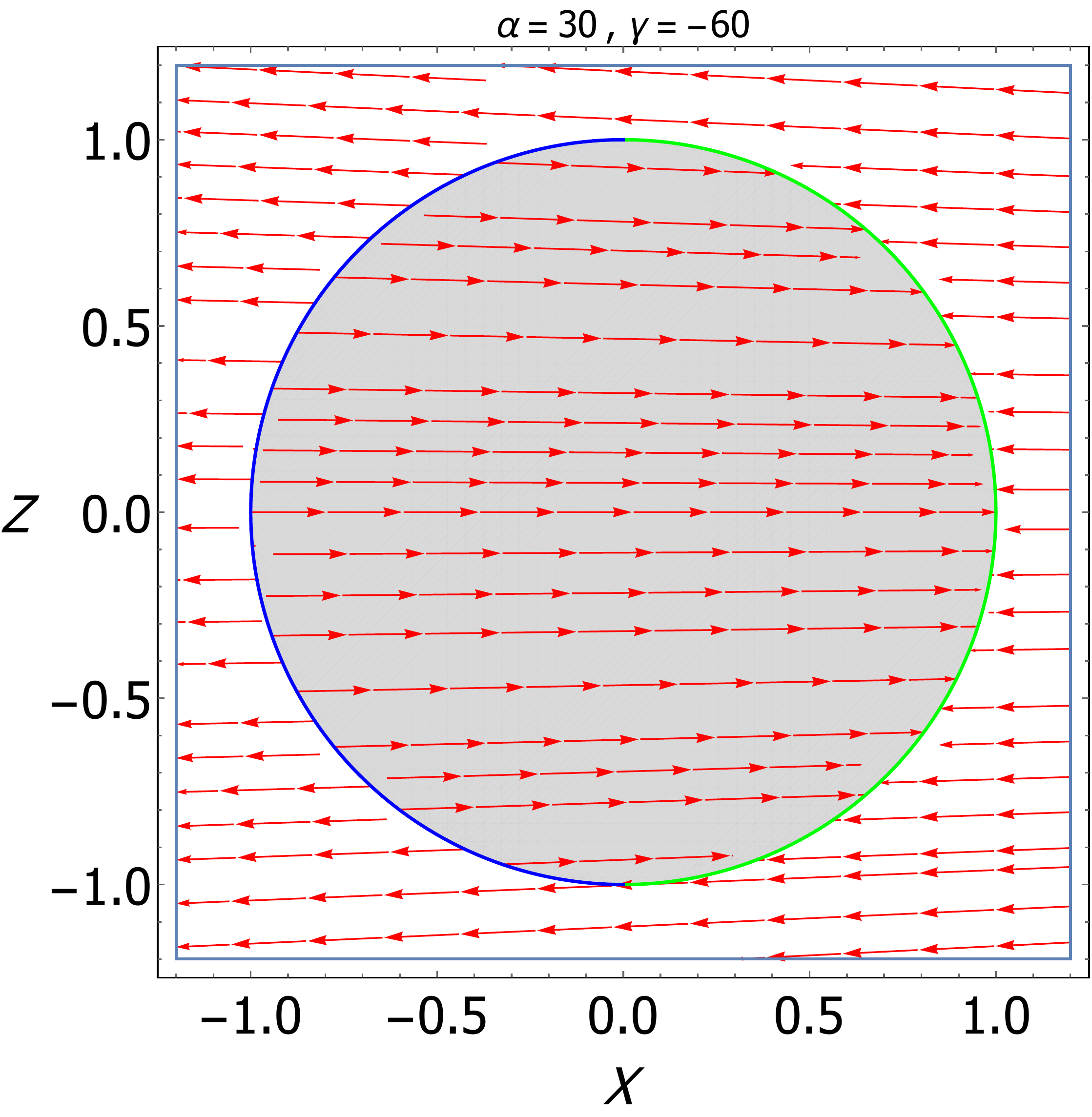}\label{fig:stable_B}}
	\qquad
	\subfigure[]{%
		\includegraphics[width=7cm,height=7cm]{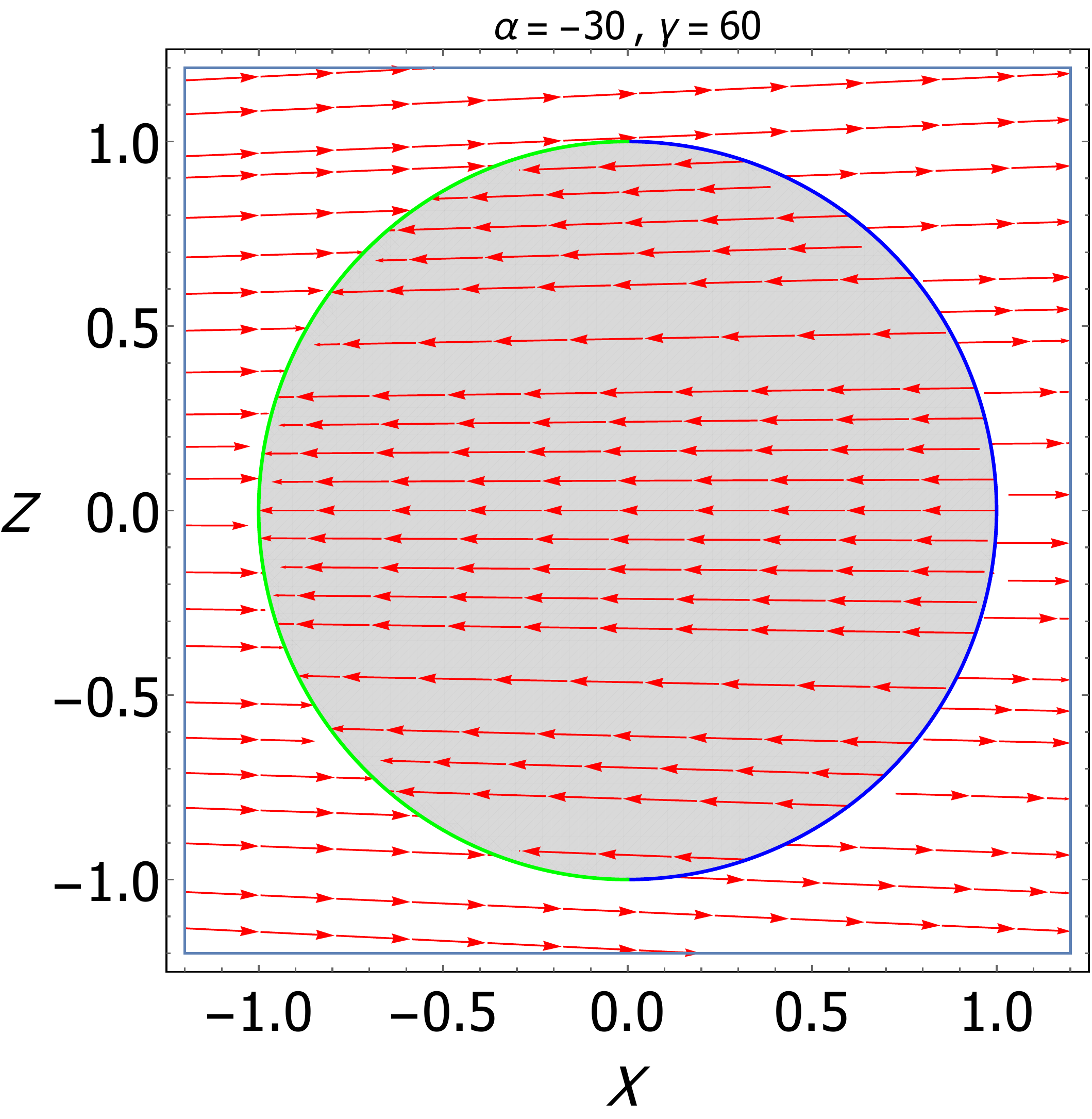}\label{fig:stable_C}}
	\caption{The figure shows the phase portrait of the autonomous system (\ref{autonomous_system}) in x-z plane for different parameter values.  In panel (a) with the parameter values $\alpha=30$ and $\gamma=-60$ the scaling solution $B$ (green colored curve) is stable attractor while the set $C$ (blue colored curve) is unstable. In panel (b), for the parameters $\alpha=-30$ and $\gamma=60$ the scaling solution $C$ (green colored curve) is stable attractor and the set of points $B$ (blue colored curve) is unstable.}
	\label{phasespace-figure_BC}
\end{figure}

It should be mentioned that in limiting case where the vector field vanishes ($\beta=0$), the modified field equations will reduce to the Einstein field equations on conventional manifold. When this vanishing displacement vector field is considered in dynamical analysis, the resulting autonomous system reduces to a two-dimensional system (with $x$ and $y$ coordinates only) and the corresponding phase space and critical points are bounded in the physical region $0\leq x^{2}+y^{2}\leq 1$ accordingly. It is worthy to note that critical points $B$ and $C$ exhibit the non-isolated sets of critical points where $z_c$ can take any real values in the  interval $[0,~ 1]$ in the frame of Lyra geometry. These points correspond to scalar field- displacement field ($\beta$) scaling solutions and they can become stable attractors within some parameter restrictions in the phase space. The stability of those sets are shown in figures (\ref{fig:stable_B}) and (\ref{fig:stable_C}) respectively. Vanishing vector field ($\beta=0$) implies immediately the dynamical variable $z$ to be zero ($z=0$) and then two critical points appear with coordinates (1,0) and (-1,0) which are analogous to that of the limiting case of sets $B$ and $C$ presented earlier (in subsection (\ref{local stability})) in 3D autonomous system. Here, the points with (1,0) and (-1,0) are completely dominated by scalar field. The corresponding figures are shown in (\ref{fig:stable_DE_2D}) and (\ref{fig:stable_FG_2D}). Therefore, the study of interacting DE-DM scenario in the framework of Lyra geometry provide some special features in the form of the sets of points $B$ and $C$.

For $\beta=0$, the 3D autonomous system (\ref{autonomous_system}) reduces to 2D autonomous system as follow:
\begin{eqnarray}
	\begin{split}
		\frac{dx}{dN}& =\frac{3}{2} \left[ -x \left(1-x^2+y^2\right)-\sqrt{\frac{2}{3}} \gamma  y^2+\sqrt{\frac{2}{3}} \alpha  \left(1-x^2-y^2\right)\right]  ,& \\
		\frac{dy}{dN}& =\frac{3}{2} y \left[1+x^2-y^2+\sqrt{\frac{2}{3}} \gamma  x\right] 
		&~~\label{autonomous_system_2D}
	\end{split}
\end{eqnarray}
The critical points and their stability arising from above 2D system are discussed below:		 
\begin{itemize}
		\item 
Critical Point : $ A_{0}=(\sqrt{\frac{2}{3}} \alpha ,0)$ appears as the point $A$ in 3D system. The existence and physical parameters are same as point $A$ given in Table \ref{physical_parameters}.  Eigenvalues of linearized Jacobian matrix of the point $A_{0}$ are: $\left\{\lambda_{1(A_0)}=\gamma \alpha+\frac{3}{2}+\alpha^2,~~\lambda_{2(A_{0})}=\alpha^2 -\frac{3}{2} \right\}$. Stablity of the point $A_0$ is same as for $A$. 

\item	
Critical Point : $ B_{0}=(1,0)$ always exists in the phase plane ($x-y$). The physical parameters for critical point $B_{0}$ are:\\ $\left\lbrace \Omega_{m}=0,\Omega_{\phi}=1,\omega^{(m)}_{eff}=\sqrt{\frac{2}{3}}\alpha,\omega^{(\phi)}_{eff}=1,\omega_{eff}=1,q=2 \right\rbrace $.
Eigenvalues of linearized Jacobian matrix for the point $B_{0}$ are: $\left\{\lambda_{1(B_{0})}=3+\frac{\sqrt{6}\gamma}{2},~~\lambda_{2(B_{0})}=3-\sqrt{6}\alpha \right\}$. Thus, the condition for the point to be stable is $ \alpha >\sqrt{\frac{3}{2}} ~\mbox{and}~ \gamma <-\sqrt{6} $.	

\item
Critical Point : $C_{0}=(-1,0)$ exists always. The physical parameters for critical point $C_{0}$ are:\\ $\left\lbrace \Omega_{m}=0,\Omega_{\phi}=1,\omega^{(m)}_{eff}=-\sqrt{\frac{2}{3}}\alpha,\omega^{(\phi)}_{eff}=1,\omega_{eff}=1,q=2 \right\rbrace $.
Eigenvalues of linearized Jacobian matrix for the point $C_{0}$ are: $\left\{\lambda_{1(C_{0})}=3-\frac{\sqrt{6}\gamma}{2},~~\lambda_{2(C_{0})}=3+\sqrt{6}\alpha \right\}$. Thus, the condition for  stablity is $ \alpha <-\sqrt{\frac{3}{2}}~\mbox{and}~ \gamma >\sqrt{6} $.

\item
Critical Point : $ D_{0}=\left(-\frac{\gamma }{\sqrt{6}},\sqrt{1-\frac{\gamma ^2}{6}} \right)$ and Critical Point : $ E_{0}=\left(-\frac{\gamma }{\sqrt{6}},-\sqrt{1-\frac{\gamma ^2}{6}} \right)$	are same as the points $D$ and $E$ in 3D system. The existence and physical parameters are same as of the points $D$ and $F$ which are shown in Table \ref{physical_parameters}. 
Eigenvalues of linearized Jacobian matrix for both the points $D_{0}$, $E_{0}$ are: $\left\{\lambda_{1(D_{0},E_{0})}=\gamma \alpha+\gamma^2 -3
,~~\lambda_{2(D_{0},E_{0})}=\frac{\gamma^2}{2}-3  \right\}$. All physical characterstics of these points are same as of the points $D$ and $E$.

\item
Critical Point : $ F_{0}=\left(-\frac{\sqrt{6}}{2(\alpha +\gamma)},\frac{\sqrt{ \alpha  (\alpha +\gamma )+\frac{3}{2}}}{ (\alpha +\gamma )} \right)$ and Critical Point : $ G_{0}=\left(-\frac{\sqrt{6}}{2(\alpha +\gamma)},-\frac{\sqrt{ \alpha  (\alpha +\gamma )+\frac{3}{2}}}{ (\alpha +\gamma )}\right)$ are same in all respect. The existence and physical parameters are same as $F$ and $G$ as provided in Table \ref{physical_parameters}. Eigenvalues of linearized Jacobian matrix for both the points $F_{0}$ and $G_{0}$ are:

$\lambda_{1(F_{0},G_{0})}=\frac{-6\alpha-3\gamma+\sqrt{-48\alpha \gamma^3 -(96\alpha^2 +63)\gamma^2 -(48\alpha^3-108\alpha)\gamma+180\alpha^2 +216}}{4\gamma+4\alpha}$,~and 
$\lambda_{2(F_{0},G_{0})}= \frac{-6\alpha-3\gamma-\sqrt{-48\alpha \gamma^3 -(96\alpha^2 +63)\gamma^2 -(48\alpha^3-108\alpha)\gamma+180\alpha^2 +216}}{4\gamma+4\alpha} $. The stability is same as for $F$ and $G$.

\end{itemize}	
Therefore, we can conclude that the analysis by assuming the limiting case of vector field $\beta=0$ gives the dynamics of interacting quintessece with varying-mass DM particles in a conventional manifold. At the same time we can mention that there is some special kind of nature after including the Lyra's geometry in the background dynamics of the model considered here. The behavior of the sets of points $B$ and $C$ can only be achieved after including Lyra geometry.

\subsection{Classical stability for the critical points :}
In cosmological perturbation theory, the speed of sound $(C_{s})$ arises naturally and it has a crucial role in studying the classical stability as well as causality of the model. In fact, $C_{s}^{2}$ appears as a coefficient of the term $\frac{k^{2}}{a^{2}}$ ($k$ is the co-moving momentum and $a$ is the usual scale factor). The sound speed can be defined as\cite{Piazza,Mahata,S.Kr.Biswas2015a,S.Kr.Biswas2015b,Garriga,Sudipta Das}
\begin{equation}\label{perturbation}
	C_{s}^{2}=\frac{\partial p_{\phi}  }{\partial \rho_{\phi}}
\end{equation}
The classical fluctuations may be considered stable when $C_{s}^{2}$ takes positive values ($C_{s}^{2}\geq 0$)  and it violates the causality for $C_{s}^{2}>1$, {\it i.e.}, when the speed of sound diverges. On the other hand, ghost instabilities may occur for $C_{s}^{2}<0$. Here, we shall investigate the classical stability of the model, avoiding the ghost instabilities. We shall now find out the term $\frac{\partial p_{\phi}  }{\partial \rho_{\phi}} $ in Eqn. (\ref{perturbation}) for the present cosmological model in terms of dynamical variables and model parameters. First, eliminating the kinetic term $\frac{1}{2} \dot{\phi}^{2}$ from the Eqns. (\ref{energy density scalar}) and (\ref{pressure scalar}), we have
\begin{equation}
p_{\phi}=\rho_{\phi}-2V(\phi)
\end{equation}
Then after differentiating, we obtain the following
\begin{equation}
	C_{s}^{2}=\frac{\partial p_{\phi}  }{\partial \rho_{\phi}}=1-\frac{2 V^{'}}{\ddot{\phi}+V^{'}}
\end{equation}
Now, by the evolution equation of scalar field in (\ref{EvolutionEqn-scalar}) and the exponential potential and exponential mass dependence in (\ref{exponential dependence}), above equation takes the form:  
\begin{equation}
	C_{s}^{2}=1+\frac{2 \gamma V}{3H \dot{\phi}-\alpha \rho_{m}}
\end{equation}
which gives the final form of $C_{s}^{2}$ (after dividing by $3H^{2}$ in both numerator and denominator of fractional term of above equation) in terms of dynamical variables (in (\ref{dynamical_variables}) ) and free parameters as
\begin{equation}
C_{s}^{2}=1+\frac{2\gamma y^{2}}{\sqrt{6} x-\alpha \left(1-x^{2}-y^{2}-z^{2} \right) }.
\end{equation}
So, from the above one obtains the required condition for classical stability:
\begin{equation}\label{classical_inequality}
\sqrt{6} x-\alpha \left(1-x^{2}-y^{2}-z^{2} \right) +2\gamma y^{2}\geq 0.
\end{equation}
The inequality in (\ref{classical_inequality}) demands the classical stability criteria for the present model. We shall now
discuss the stability of the model at each critical point where $x$, $y$ and $z$ are the coordinates of corresponding critical point. We observe that
\begin{itemize}
\item
The critical point $A$ corresponds to classical stability for~ $0\leq \alpha \leq \sqrt{\frac{3}{2}}.$\
\item
For  $0\leq z_{c}<1$ the set of critical points $B$ is stable classically whereas $C$ is unstable there. But, both of them are classical stable (limiting) when $z_{c}=1.$\
\item
Conditions for classical stability of the critical points $D$ and $E$ are:\\
(i)~$-\sqrt{6}\leq \gamma \leq -\sqrt{3},$ ~or (ii)~$ 0\leq \gamma \leq \sqrt{3}.$\
\item
Finally, the points $F$ and $G$ are stable classically for the following parameter restrictions:\\
(i)~$\alpha =0~~\mbox{and}~~ \left(\gamma \leq -\sqrt{3}~~\mbox{or}~~\gamma \geq \sqrt{3}\right)$ ,~or\\
(ii)~$0<\alpha <\sqrt{\frac{3}{2}}~~\mbox{and}~~ \left(\frac{-2 \alpha ^2-3}{2 \alpha }\leq \gamma \leq -\frac{\sqrt{\alpha ^2+12}}{2}-\frac{\alpha }{2}~~\mbox{or}~~ \gamma \geq \frac{\sqrt{\alpha ^2+12}}{2}-\frac{\alpha }{2}\right)$ ,~or\\
(iii)~$\alpha =\sqrt{\frac{3}{2}}~~\mbox{and}~~ \left(\gamma =-\sqrt{6}~~\mbox{or}~~ \gamma \geq \sqrt{\frac{3}{2}}\right)$ ,~or\\
(iv)~$\alpha >\sqrt{\frac{3}{2}}~~\mbox{and}~~ \gamma \geq \frac{\sqrt{\alpha ^2+12}}{2}-\frac{\alpha }{2}.$
\end{itemize}
Therefore, from the study of local stability as well as classical stability one can draw the following conclusion:
Critical points may be stable locally, but not classically.
Critical points may be stable classically, but not locally.
Critical points stable locally as well as classically.
Critical point is not stable locally as well as classically.
However, among all of the above possibilities, we like to discuss only one possible region in which the ``critical points are stable locally as well as classically" and this is the acceptable and promising one in cosmological study. So, our objective is to obtain the restrictions on model parameters for which the criteria are satisfied.  
\begin{itemize}
\item
The critical point $A$ is stable locally as well as classically for $0<\alpha <\sqrt{\frac{3}{2}}~~\mbox{and}~~ \gamma <\frac{-2 \alpha ^2-3}{2 \alpha }.$\
\item
The set of critical points $B$ is stable locally as well as classically for 
$0\leq z_{c}<1~~\mbox{and}~~ \alpha >\sqrt{\frac{3}{2}} \sqrt{\frac{1}{1-z_{c}^{2}}}~~\mbox{and}~~ \gamma <-\sqrt{6} \sqrt{\frac{1}{1-z_{c}^{2}}}.$\
\item
The set of critical points $C$ is stable locally for the restrictions: 
$0\leq z_{c}<1~~\mbox{and}~~ \alpha <-\sqrt{\frac{3}{2}} \sqrt{\frac{1}{1-z_{c}^{2}}}~~\mbox{and}~~ \gamma >\sqrt{6} \sqrt{\frac{1}{1-z_{c}^{2}}}$, but not stable classically for that restrictions. Note that set $C$ is stable classically (limiting) only for $z_{c}=1.$\
\item
The points $D$ and $E$ are stable locally as well as classically when the following conditions are satisfied:\\
(i)~$\left( \alpha <0~~\mbox{and}~~ 0\leq \gamma \leq \sqrt{3}\right) $ ,~or\\
(ii)~$ 0\leq \alpha \leq \sqrt{\frac{3}{2}}~~\mbox{and}~~ \left(-\frac{\sqrt{\alpha ^2+12}}{2}-\frac{\alpha }{2}<\gamma \leq -\sqrt{3}~~\mbox{or}~~ 0\leq \gamma <\frac{\sqrt{\alpha ^2+12}}{2}-\frac{\alpha }{2}\right) $ ,~or\\
(iii)~$ \alpha >\sqrt{\frac{3}{2}}~~\mbox{and}~~ \left(-\sqrt{6}<\gamma \leq -\sqrt{3}~~\mbox{or}~~ 0\leq \gamma <\frac{\sqrt{\alpha ^2+12}}{2}-\frac{\alpha }{2}\right) .$\
\item
The equilibrium points $F$ and $G$ are locally as well as classically stable for:\\
(i)~$ \alpha =0~~\mbox{and}~~ \left(\gamma <-\sqrt{3}~~\mbox{or}~~ \gamma >\sqrt{3}\right) $ ,~or\\
(ii)~$ 0<\alpha <\sqrt{\frac{3}{2}}~~\mbox{and}~~ \left(\frac{-2 \alpha ^2-3}{2 \alpha }<\gamma <-\frac{\sqrt{\alpha ^2+12}}{2}-\frac{\alpha }{2}~~\mbox{or}~~ \gamma >\frac{\sqrt{\alpha ^2+12}}{2}-\frac{\alpha }{2}\right) $ ,~or\\
(iii)~$\alpha \geq \sqrt{\frac{3}{2}}~~\mbox{and}~~ \gamma >\frac{\sqrt{\alpha ^2+12}}{2}-\frac{\alpha }{2}.$ 
\end{itemize} 
  

	
\begin{table}[tbp] \centering
\caption{The Critical Points and the corresponding physical parameters  are presented.}%
\begin{tabular}
[c]{cccccccc}\hline\hline
\textbf{Critical Points}&$\mathbf{\Omega_{m}}$& $\mathbf{\Omega_{\phi}}$ & $\mathbf{\Omega_{\beta}}$ &
 $\mathbf{\omega^{(\phi)}_{eff}}$ & $\mathbf{\omega_{eff}}$ & $q$ &
\\\hline
$A  $ & $1-\frac{2 \alpha ^2}{3}$ & $\frac{2 \alpha ^2}{3}$ &
$0$ & $\frac{2 \alpha ^2}{3}$ & $\frac{2 \alpha ^2}{3}$ & $\alpha ^2+\frac{1}{2}$ \\
$B  $ & $0$ & $1-z_{c}^{2}$ &
$z_{c}^{2}$ & $1$ & $1$ & $2$\\
$C  $ & $0$ & $1-z_{c}^{2}$ &
$z_{c}^{2}$ & $1$ & $1$ & $2$ \\
$D $ & $0$ & $1$ &
$0$ & $\frac{\gamma ^2}{3}-1$ & $\frac{\gamma ^2}{3}-1$ & $\frac{{\gamma}^{2}}{2}-1$\\
$E  $ &  $0$ & $1$ &
$0$ & $\frac{\gamma ^2}{3}-1$ & $\frac{\gamma ^2}{3}-1$ & $\frac{{\gamma}^{2}}{2}-1$\\
$F  $ & $\frac{\gamma  (\alpha +\gamma )-3}{(\alpha +\gamma )^2}$ & $\frac{\alpha  (\alpha +\gamma )+3}{(\alpha +\gamma )^2}$ & $0$ & $-\frac{\alpha }{\alpha +\gamma }$ & $-\frac{\alpha }{\alpha +\gamma }$ & $\frac{\gamma -2 \alpha }{2 (\alpha +\gamma )}$ \\
$G  $ & $\frac{\gamma  (\alpha +\gamma )-3}{(\alpha +\gamma )^2}$ & $\frac{\alpha  (\alpha +\gamma )+3}{(\alpha +\gamma )^2}$ & $0$ & $-\frac{\alpha }{\alpha +\gamma }$ & $-\frac{\alpha }{\alpha +\gamma }$ & $\frac{\gamma -2 \alpha }{2 (\alpha +\gamma )}$ \\

 \\\hline\hline
\end{tabular}
\label{physical_parameters} \\

\end{table}%
%
\begin{figure}
	\centering
	\subfigure[]{%
		\includegraphics[width=7cm,height=7cm]{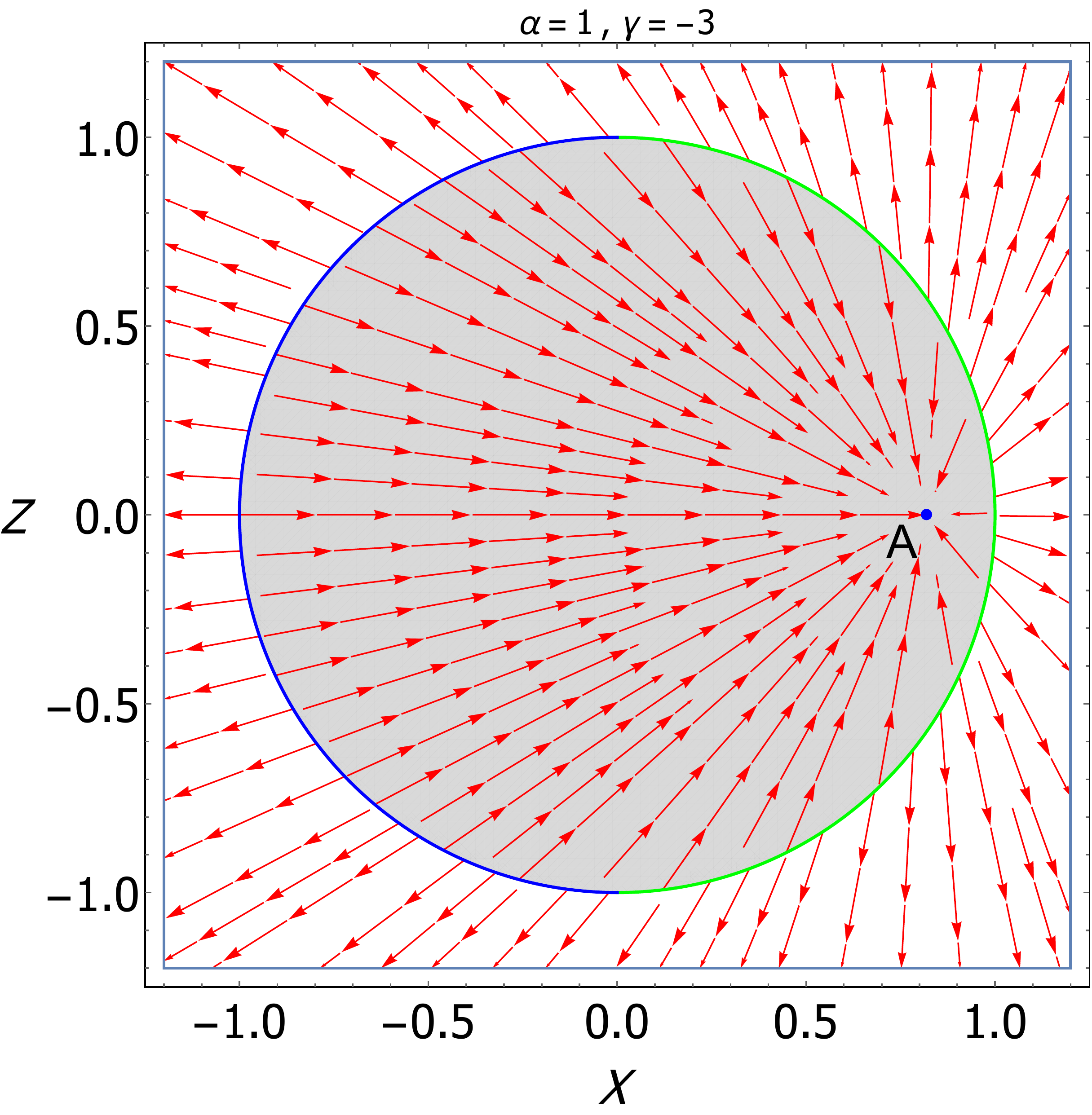}\label{fig:stability_A}}
	\qquad
	\subfigure[]{%
		\includegraphics[width=7cm,height=7cm]{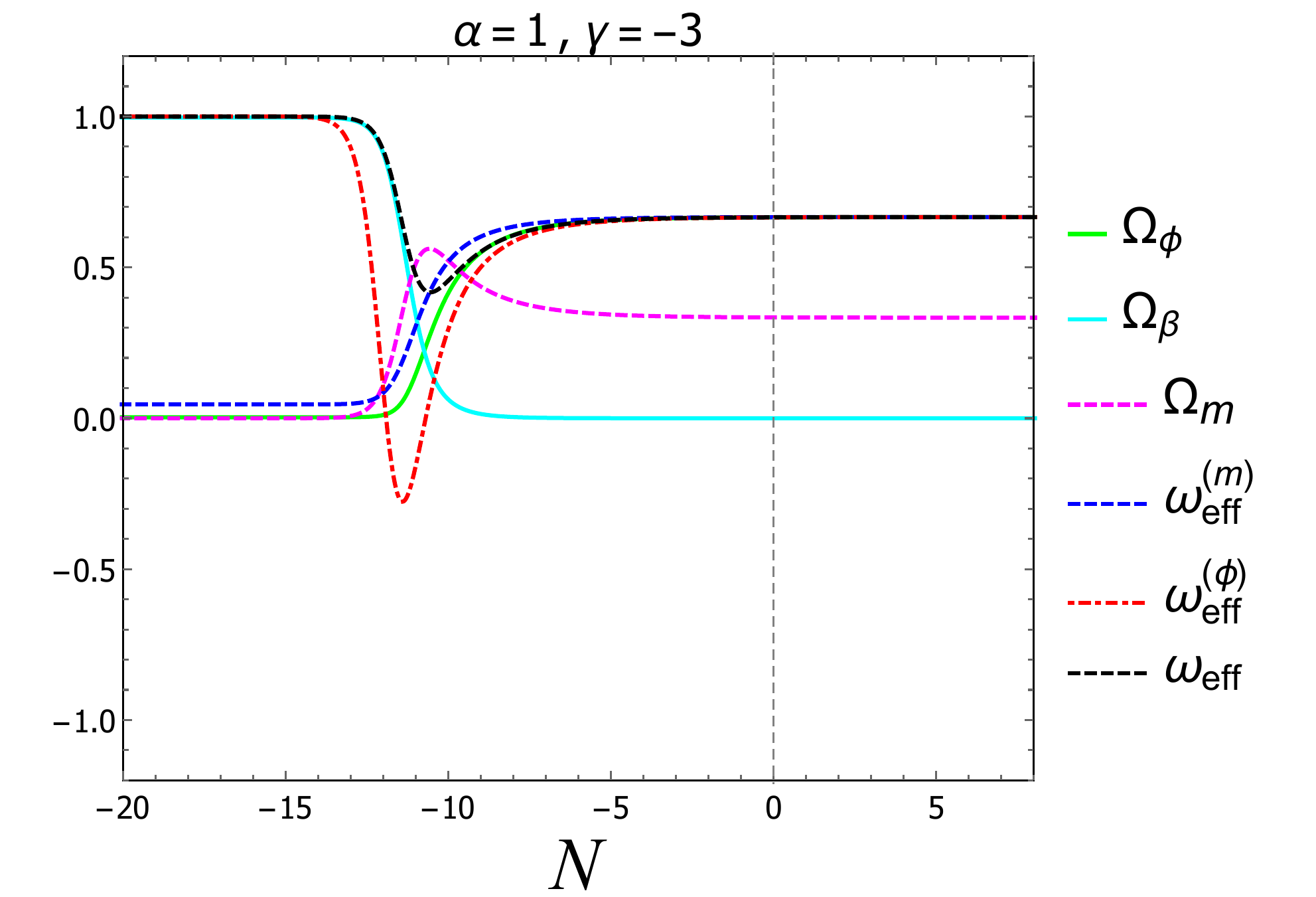}\label{fig:evolution_A}}
	\caption{The figure is plotted with $\alpha=1$ and $\gamma=-3$.  Panel (a) shows the phase projection on $x-z$ plane, where the scaling solution A is a stable attractor. While panel (b) shows the time evolution of the cosmological parameters with initial conditions: $ x[0]=0.816,~ y[0]=0.001,~ z[0]=0.001$ where late time scaling attractor approaches in decelerated era.}
	\label{phasespace-figure_A}
\end{figure}
\section{Cosmological Implications}\label{cosmological implications}

In the context of Lyra's geometry, interacting quintessence with varying-mass dark matter is investigated where potential of scalar field and mass of dark matter particles are considered to vary with exponential form of scalar field $\phi$. Dynamical systems tools have been applied to understand the dynamics of the cosmological model undertaken. Critical points are analyzed by evaluating eigenvalues of linearized Jacobian matrix at critical points. Classical stability are also performed to study the model stability. We have obtained parameter restrictions in which the points are stable locally as well as classically. This is reported in previous section. From the dynamical system perspective, the present model exhibits some cosmological viable scenarios which are the following:  \\

\paragraph*{{\Large (a) Scalar field dominated attractor solutions:}}

\begin{figure}
	\centering
	\subfigure[]{%
		\includegraphics[width=7cm,height=7cm]{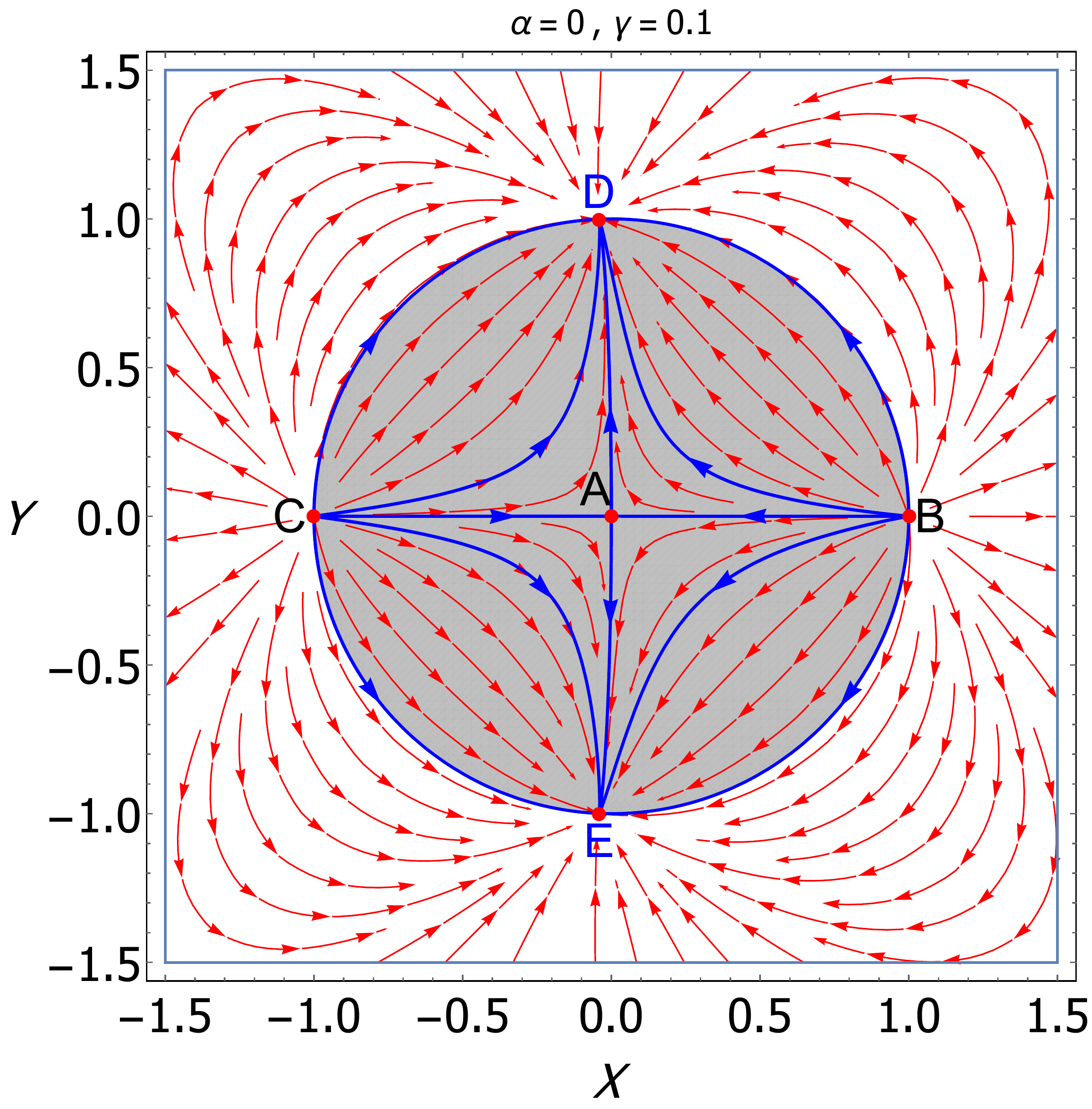}\label{fig:stable_DE_2D}}
	\qquad
	\subfigure[]{%
		\includegraphics[width=9cm,height=7cm]{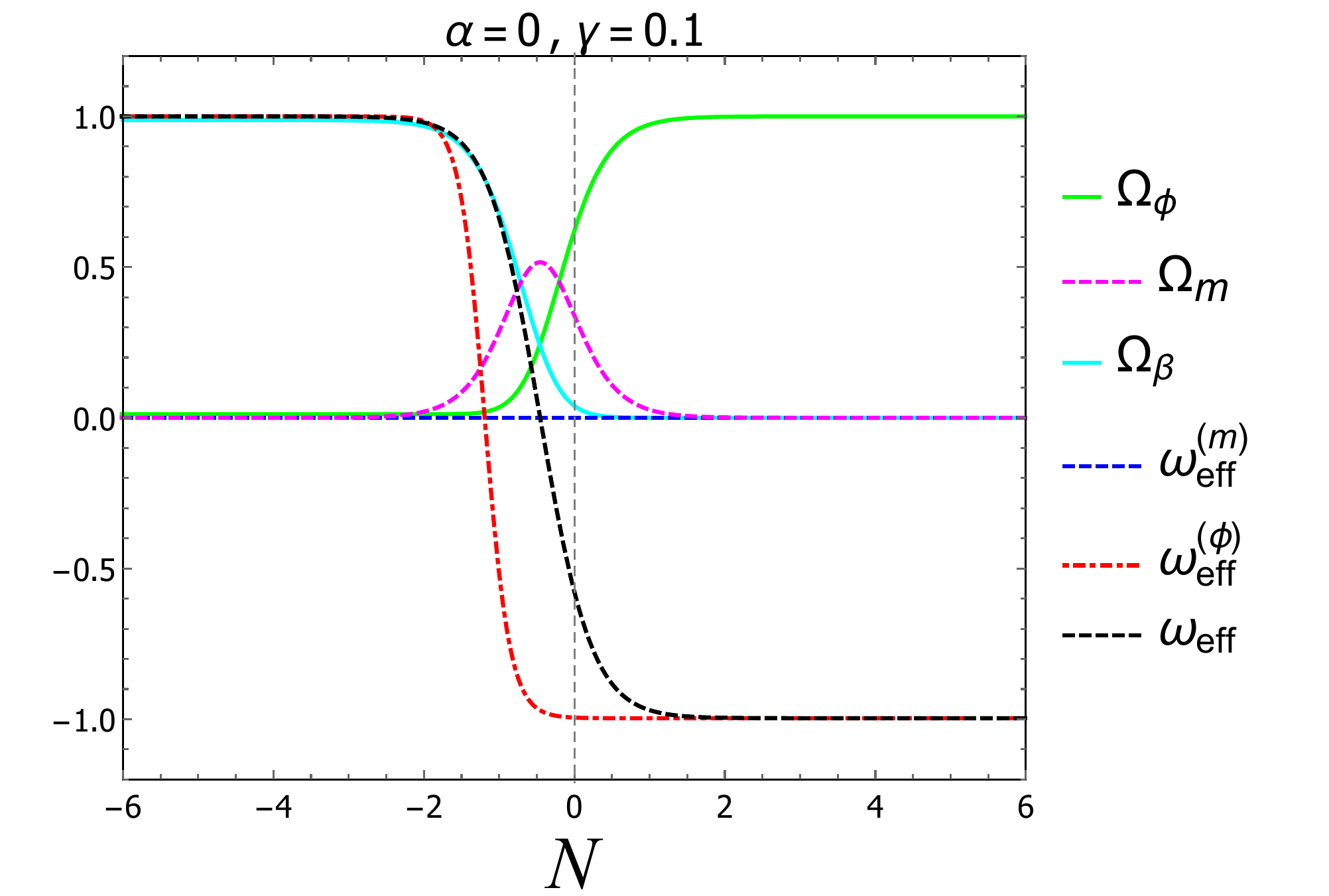}\label{fig:evolution_DE}}
		\qquad
		\subfigure[]{%
	\includegraphics[width=7cm,height=7cm]{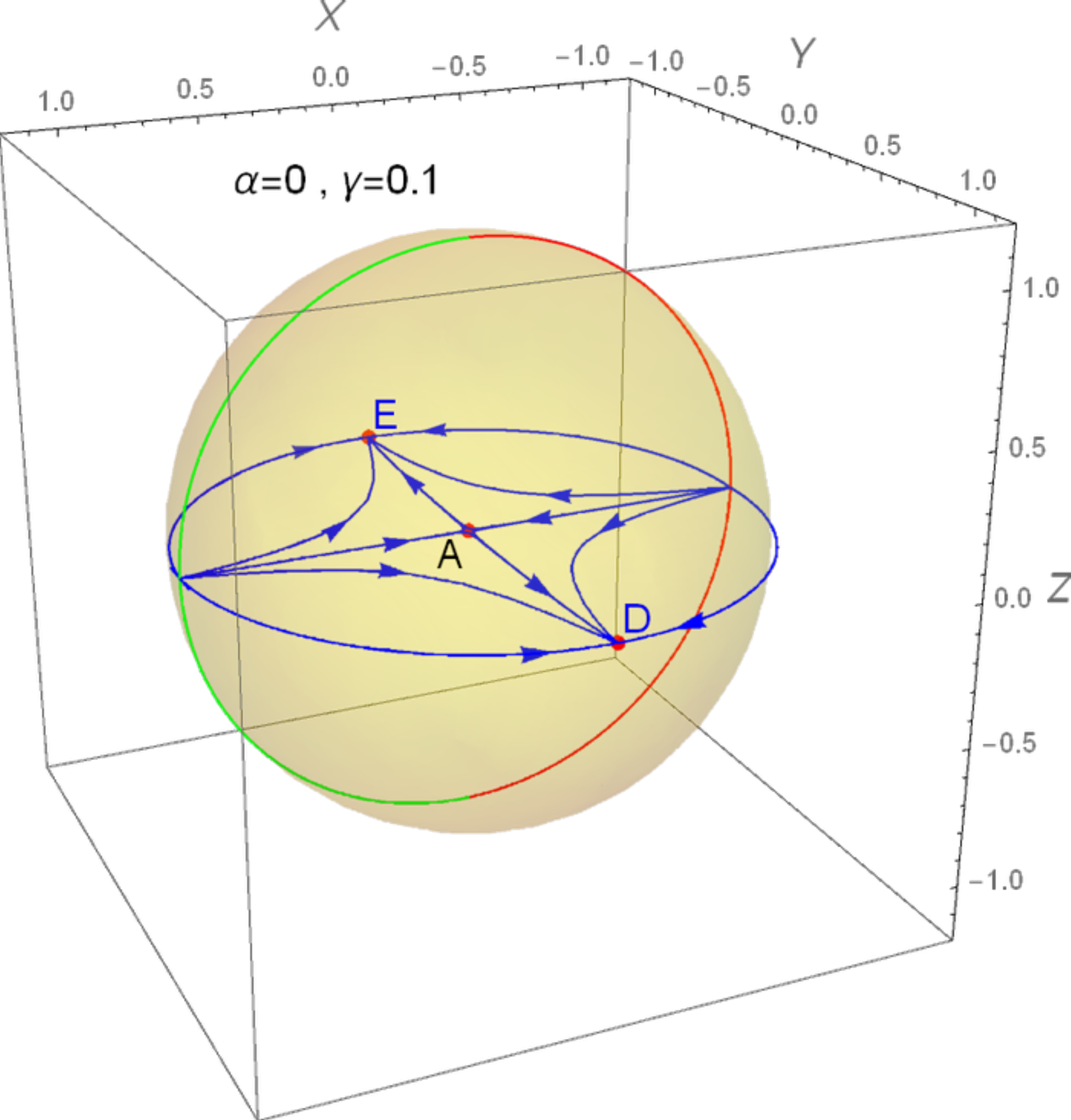}\label{fig:stable_DE_3D}}
	\caption{The figure is plotted for parameters $\alpha=0, \gamma=0.1$. Panel (a) shows the phase space projection on $x-y$ plane where the scalar field dominated solutions $D$ and $E$ are stable attractors. In panel (b), with the initial conditions: $x[0]= -0.041,~
		y[0]=0.787,~  z[0]=0.2 $, the time evolution of physical parameters show that the late time attractor is approaching in accelerated era with cosmological constant. Finally, 3D plot in panel (c)  shows that the set of points (an arc of circle) with green colored and red colored arc denoting $B$ and $C$ respectively are source, whereas the point $A$ is saddle and $D$ and $E$ are late time attractors.}
	\label{phasespace-figure_DE}
\end{figure}

Scalar field dominated late time attractor solutions are represented by the critical point $D$ and $E$ which are accelerating in nature in the phase space. Depending upon some restrictions on the free parameters $\alpha$ and $\gamma$, evolution of the universe is attracted in quintessence era for non-constant (i.e., for $\gamma\neq 0$) potential of scalar field. On the other hand, the points $D$ and $E$ correspond to late time accelerated de Sitter solutions for constant potential of scalar field ($\gamma=0$). In this case, the evolution of de Sitter universe is attracted by cosmological constant ($\Omega_{\phi}=1, \omega_{eff}=q=-1$). Note that the critical points cannot solve the coincidence problem since $\Omega_{\phi}=1$ for those points is satisfied. Figure (\ref{phasespace-figure_DE}) shows the phase portrait of the system (\ref{autonomous_system}) for parameter values $\alpha=0$ and $\gamma=0.1$. In this figure, the subfigures  \ref{fig:stable_DE_2D} and \ref{fig:stable_DE_3D} exhibit that the points $D$ and $E$ are DE dominated stable attractors in late times and also these are connected through a matter dominated saddle point $A$. Here, the sets of points $B$ and $C$ are unstable source. Time evolution of different cosmological parameters (density parameters: $\Omega_{\phi}$, $\Omega_{m}$ and $\Omega_{\beta}$; effective equation of state parameter for DE and DM: $\omega^{(m)}_{eff}$ and $\omega^{(\phi)}_{eff}$; effective equation for the model: $\omega_{eff}$) are shown in the fig.\ref{fig:evolution_DE}, where the late time accelerated universe evolves in quintessence era and ultimately late time evolution of the universe is attracted by cosmological constant. \\

\paragraph*{{\Large (b) Scalar field-displacement field scaling attractor solutions:}}
Late time scaling attractors have generally received a lot of interest from cosmological point of view. We have found some non-isolated set of points namely, set $B$ and set $C$ which represent the scalar field DE-  displacement vector field ($\beta$) scaling solutions in the phase space. Each set is non-hyperbolic in nature. Each of these sets having exactly one vanishing eigenvalue is called normally hyperbolic set. From the stability analysis, we have found that the sets $B$ and $C$ can be late time stable attractor solutions with some parameter restrictions in the phase space. Expansion of the universe near the sets is always decelerated ($\omega_{eff}=1, q=2$). In particular, $B$ for $z_c=0$ (a specific point $B(1,0,0)$) representing scalar field dominated ($\Omega_{\phi}=1$) decelerated universe, is a future attractor for $\alpha>\sqrt{\frac{3}{2}}$, $\gamma<-\sqrt{6}$, otherwise it is a source. Further, the set $C$ represents decelerated stable attractor for  $\alpha<\sqrt{\frac{3}{2}}$, $\gamma>\sqrt{6}$, otherwise it is a source.

On the other hand, for $z_c=1$, displacement field dominated ($\Omega_{\beta}=1$) solutions $B$ and $C$ will coincide to each other and will represent the decelerated source in the phase space. In figure (\ref{phasespace-figure_BC}) the projection of phase portrait on $x-z$ plane shows that the sets $B$ and $C$ are stable attractors for different values of parameters involved. For $\alpha=30$ and $\gamma=-60$, the fig.\ref{fig:stable_B} shows that the DE-displacement field ($\beta$) scaling solution $B$ in green colored arc is stable attractor while the set $C$ in blue colored arc is unstable source. On the other hand, fig.\ref{fig:stable_C} shows that the scaling solution $C$ is stable attractor and set $B$ is unstable source for $\alpha=-30$ and $\gamma=60$ . \\

\paragraph*{{\Large (c) Scalar field- dark matter scaling attractor solutions:}}  
Dark matter-dark energy scaling solution described by the point $A$ corresponds to a decelerated ($q=\frac{1}{2}+\alpha^{2}$) universe in its cosmic evolution. For the parameter $\alpha\longrightarrow \pm \sqrt{\frac{3}{2}}$, DE mimicks stiff fluid ($\omega_{\phi}=1$) and dominates over the DM. On the other hand, for $\alpha=0$, the critical point $A$ will become completely DM (dust) dominated solution and saddle like in nature. And it then describes the intermediate dust dominated decelerated phase of the universe ($\Omega_{m}=1, \Omega_{\phi}=1, \omega_{eff}=0, q=\frac{1}{2}$) which is an essential part to show the complete evolution in an interacting scenario. Interestingly, for non-constant potential (i.e., for $\alpha\neq 0$), the point can describe a late time stable attractor with $0<\Omega_{\phi}<1$ and $q=\alpha^{2}+\frac{1}{2}$ which correspond to a decelerated universe though it cannot solve the coincidence problem. Figure (\ref{phasespace-figure_A}) shows the projection of phase portrait on $x-z$ plane and the evolution of cosmological parameters for $\alpha=1$ and $\gamma=-3$ of the system (\ref{autonomous_system}). Fig.\ref{fig:stability_A} shows that the point $A$ describes the late time attractor. In fig.\ref{fig:evolution_A}, the time evolution of cosmological parameters show that the universe is attracted in decelerated era.   \\

\begin{figure}
	\centering
	\subfigure[]{%
		\includegraphics[width=7cm,height=7cm]{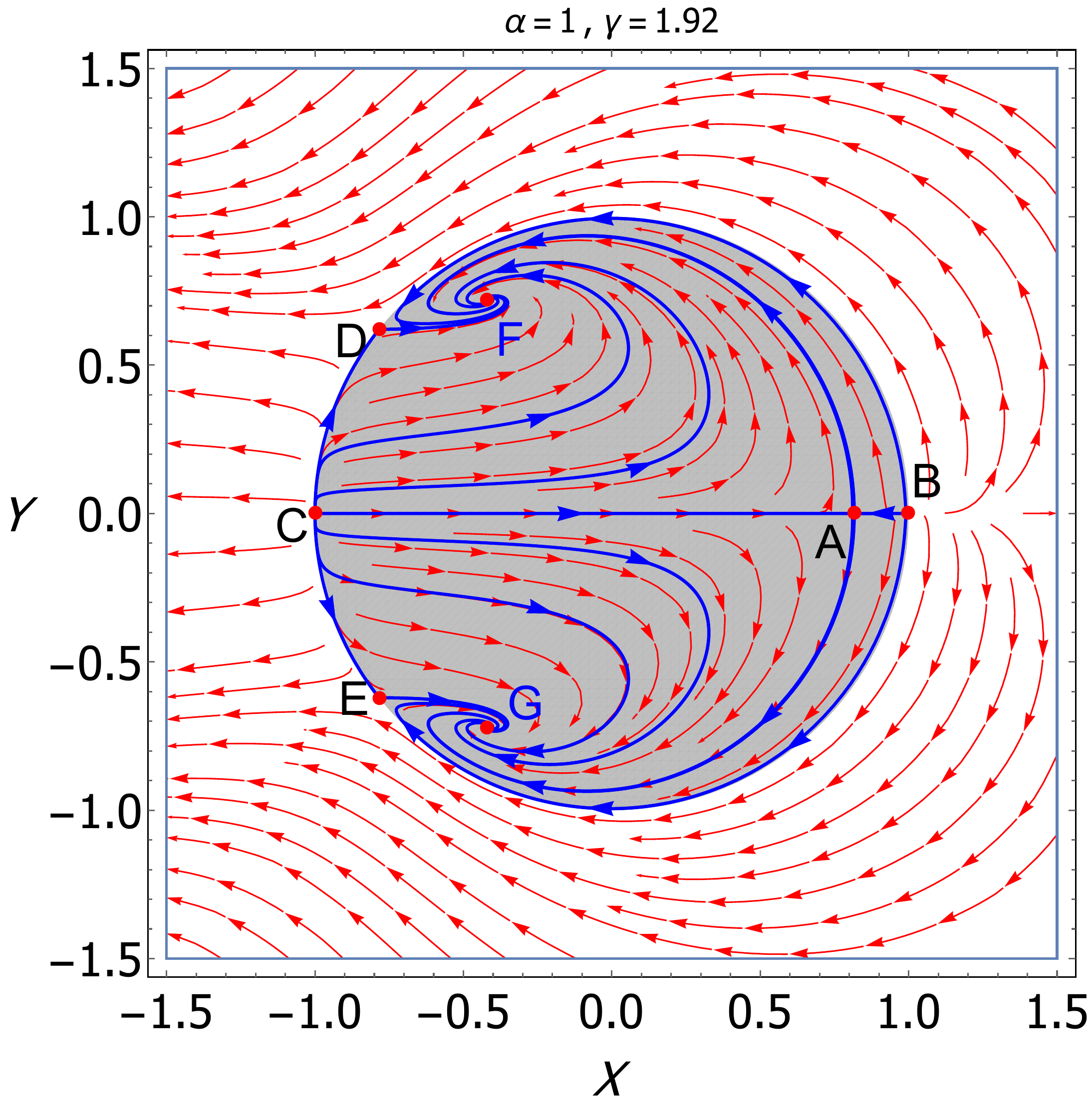}\label{fig:stable_FG_2D}}
	\qquad
	\subfigure[]{%
		\includegraphics[width=9cm,height=7cm]{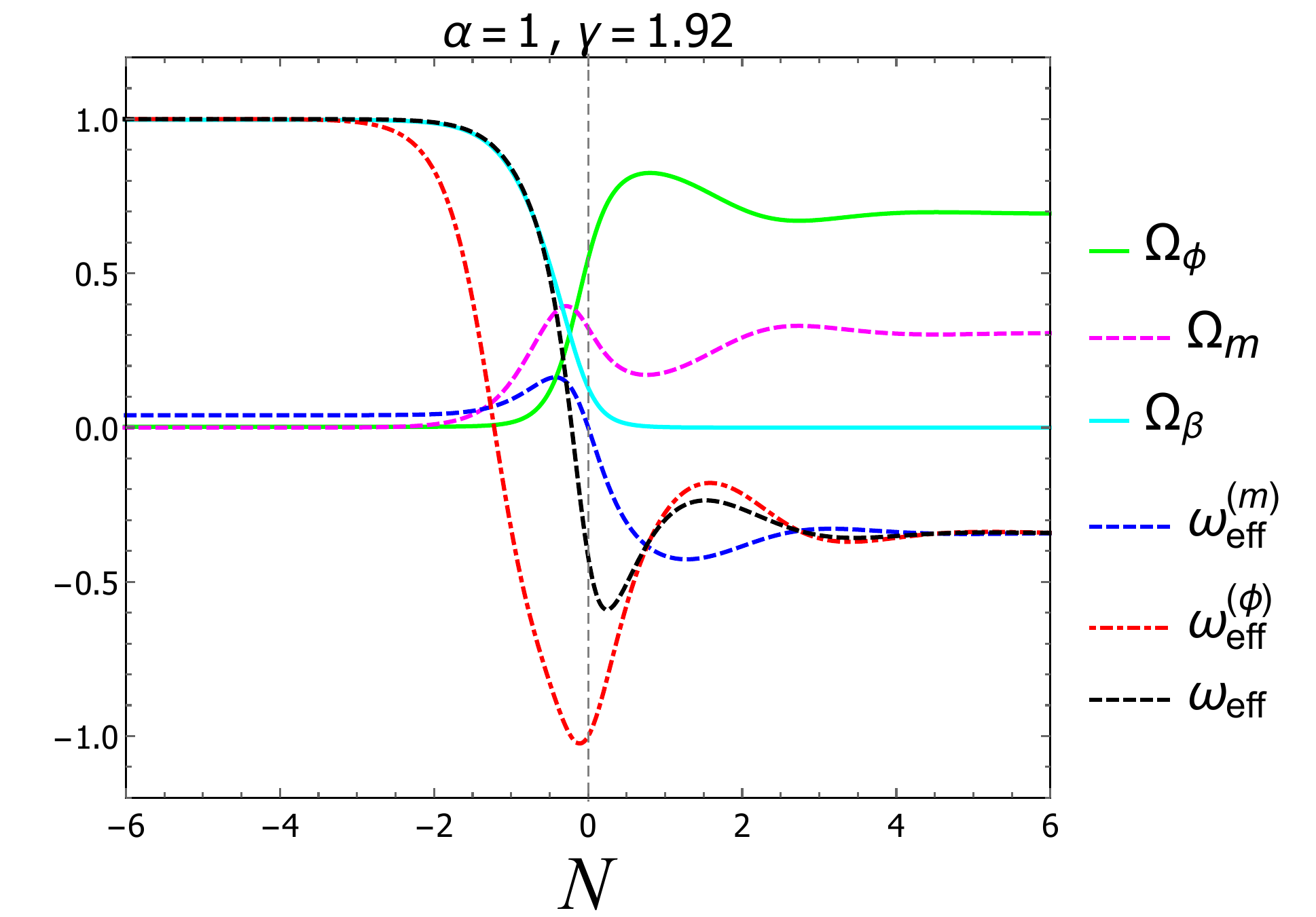}\label{fig:evolution_FG}}
		\qquad
		\subfigure[]{%
	\includegraphics[width=7cm,height=7cm]{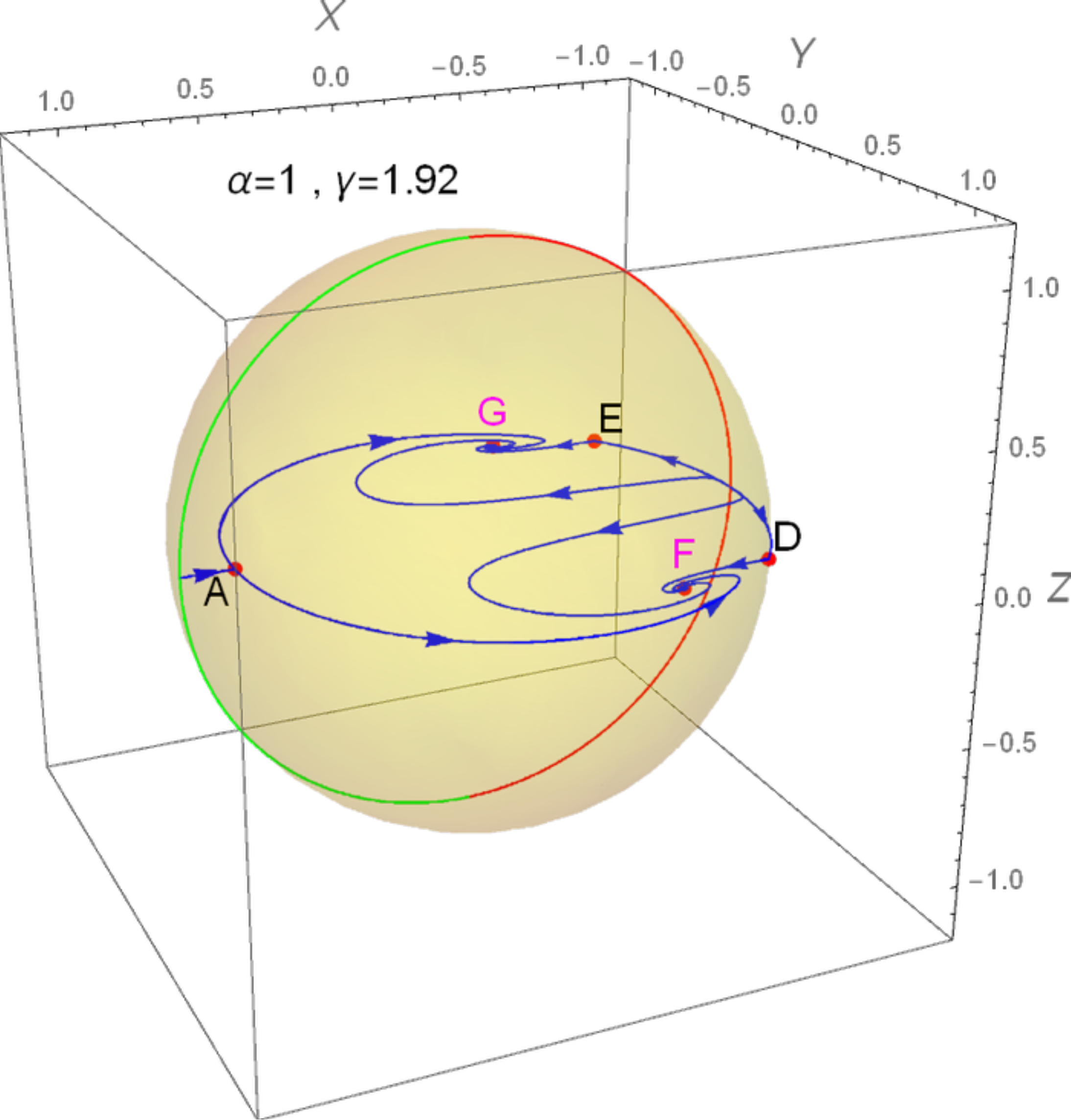}\label{fig:stable_FG_3D}}
	\caption{The figure shows the phase portrait of the autonomous system (\ref{autonomous_system}) for parameter values $\alpha=1$ and $\gamma=1.92$. Panel (a) shows that the phase projection on $x-y$ plane where critical points $F$ and $G$ represent the late time scaling attractor. In panel (b), the evolution of physical parameters with initial conditions $x[0]=0.001,~ y[0]=0.74,~ z[0]=0.36 $, show that the scaling solutions $F$ and $G$ are attracted in quintessence era. In panel (c), 3D phase space shows the scaling solutions $F$ and $G$ are late time attractors.}
	\label{phasespace-figure_FG}
\end{figure}

Finally, we have obtained late time accelerated dark matter-dark energy scaling attractor solutions namely, the critical points $F$ and $G$. Energy contribution from displacement field is absent to the total energy distribution of the universe ($\Omega_{\beta}=0$). Stability analysis confirms that the points can be solutions at late times and can provide interesting dynamics in late phase evolution of the universe with $\omega_{eff}<-\frac{1}{3}$ satisfying constant ratio of density parameters for DE and DM in the evolution of the universe as $r=\frac{\Omega_{\phi}}{\Omega_{m}}=\frac{\alpha  (\alpha +\gamma )+3}{\gamma  (\alpha +\gamma )-3}$, as a result of which the coincidence problem could be alleviated. Therefore, one can conclude that the points correspond to late time accelerated evolution of the universe attracted in quintessence era with $0<\Omega_{\phi}<1$ alleviating the coincidence problem. Fig. \ref{fig:evolution_FG} for $\alpha=1$ and $\gamma=1.92$, with proper initial conditions shows that the late-time accelerated evolution of the universe is attracted in quintessence era satisfying $\omega^{(m)}_{eff}=\omega^{(\phi)}_{eff}$ in the attractor regime solving the coincidence problem. The above condition ($\omega^{(m)}_{eff}=\omega^{(\phi)}_{eff}$) indicates the tracker behavior of the solutions. It is worthy to note that the fig. \ref{fig:stable_FG_2D} and fig.\ref{fig:stable_FG_3D} for $\alpha=1$ and $\gamma=1.92$ show that the physical parameter values: $[\Omega_{\phi} \approx 0.7,~ \Omega_{m} \approx 0.3,~ \omega_{eff} \approx -0.34,~ \omega^{(m)}_{eff} = \omega^{(\phi)}_{eff} \approx -0.34]$, can support the present evolution of the universe in quintessence era.

\section{Discussions with concluding remarks}\label{discussion}

An interacting DE scenario in the framework of Lyra geometry is investigated in the background of spatially flat FLRW universe where quintessence scalar field is taken as the model of DE and pressure-less dust is taken as DM. The mass of DM particles varies with time through a scalar field $\phi$ in the sense that decaying of DM particles reproduce the scalar field DE. Exponential potential $V(\phi)=V_{0}~ \mbox{exp}(\gamma\phi)$ and exponential mass dependence $M_{m}(\phi)=M_{0}~ \mbox{exp}(-\alpha\phi)$ are considered here. Additionally, a modification on geometric part of the Einstein's equations is also considered. In this modification (Lyra's geometry), a displacement vector field arises as a natural consequence of intrduction of a gauge function into the structureless manifold. By using the so called normal gauge condition, we obtain the modified field equation in which the displacement vector field  becomes a dynamical quantity. In this work, a time-varying displacement vector field is considered and it varies with scale factor as $\beta(t) \propto a^{-3}(t)$ in the evolution equations. For the limiting case of $\beta$, {\it i.e.}, when vector field vanishes, the modified field equations in Lyra geometry recover the Eienstein's equations as in general relativity. This case corresponds to an interacting DM and quintessence on conventional 4D manifold. In the framework of Lyra geometry, the Modified Friedmann equation, acceleration equation, conservation equation for DE, displacement field and conservation equations for varying-mass DM are obtained.  Since the governing equations are very much complicated to solve them analytically, dynamical systems tools have been undertaken to study the model qualitatively. For that purpose, we convert the equations into an autonomous system of non-linear ordinary differential equations by suitable choice of dimensionless variables which are normalized over Hubble scale. Linear stability (from Hartman-Grobman theory) for the critical points as well as classical stability (from speed of sound) for the model are investigated. From the dynamical analysis, we have found some cosmological viable scenarios as follows:\\

DM-scalar field scaling solution namely, the critical point $A$ exists for $-\sqrt{\frac{3}{2}}\leq \alpha \leq \sqrt{\frac{3}{2}}$. Depending upon $\alpha$, the point can represent late time attractor, or saddle like intermediate phase of the universe, which is always decelerating in nature. In particular, it can be scalar field dominated solution for $\alpha=\pm \sqrt{\frac{3}{2}}$ where scalar field behaves as stiff matter. On the other hand, for $\alpha=0$ the point describes a dust dominated decelerated universe and has the transient nature (since the critical point is saddle like in nature) in its evolution where scalar field behaves as dust matter. \\

From the dynamical point of view, there are two critical points $B$ and $C$ depict the similar evolutionary scheme of the universe. The points are normally hyperbolic set. From the analysis, we observe that the sets correspond to scalar field- displacement field scaling solution. For $z_c=1$, the total energy density is contributed by the displacement field $\beta(t)$ due to Lyra's manifold. We have obtained an ever decelerating universe near the sets. Depending upon some parameter restrictions, the sets can represent late time stable attractor with $\beta$ energy dominated decelerated universe which has been discussed in earlier section.
In the case of limiting $\beta=0$, we have obtained the scalar field dominated solutions $B_0$ and $C_0$ as analogous to that of $B$ and $C$.\\

Completely scalar field dominated solutions namely, critical points $D$ and $E$ exist in the parameter space $-\sqrt{6}\leq \gamma\leq \sqrt{6}$. In this interval the points are same in all respects. They exhibit the late time accelerated universe which is attracted either in quintessence era or by cosmological constant. Late time accelerated evolution of the universe is attracted by cosmological constant for $\gamma=0$. Interestingly, here the expansion of the universe is exponential and it represents the late time de Sitter solution. \\    
 
Finally, the scalar field- matter scaling solutions are achieved by the critical points $F$ and $G$. These points represent the late time attractor solutions corresponding to the accelerating universe attracted in quintessence era and it has similar energy densities of DE and DM in its evolution which can solve the coincidence problem.\\
From the classical stability of the model, we have obtained the constraints of parameters which lead to the stability of critical points locally as well as classically.

In conclusion, one can state that the study of interacting DE scenario with varying-mass DM particles in framework of Lyra geometry provides some cosmological viable results. Some of critical points exhibit scalar field dominated late time accelerated attractor solutions and some represent decelerated scalar field-displacement field scaling attractor solutions but for the both cases, coincidence problem cannot be solved. On the other hand, some critical points correspond to the accelerated scalar field-DM scaling attractor solutions which can alleviate the coincidence problem, because in the attractor regime both the scalar field and DM energy densities scale in a similar order.


\section*{Acknowledgments}
The author Goutam Mandal acknowledges UGC, Govt. of India for providing Junior Research Fellowship [Award Letter No.F.82-1/2018(SA-III)] for Ph.D. The authors are thankful to Prof. Subenoy Chakraborty for valuable comments on Lyra geometry. The authors are indebted to the reviewer for his/her constructive comments for which the quality of the paper has improved significantly.


\begin{thebibliography}{}
	\bibitem{Riess:1998cb}
	A.~G.~Riess {\it et al.}  [Supernova Search Team Collaboration],
	{\it	Astron.\ J.\ }  {\bf 116} (1998), 1009
	(astro-ph/9805201).
	
	\bibitem{Perlmutter:1998np}
	S.~Perlmutter {\it et al.}  [Supernova Cosmology Project Collaboration],
	{\it Astrophys.\ J.\ } {\bf 517} (1999), 565
	(astro-ph/9812133).
	
	
	
	\bibitem{Ade:2015xua}
	P.~A.~R.~Ade {\it et al.} [Planck Collaboration],
	{\it Astron. Astrophys.}  \textbf{594} (2016), A13
	(arXiv:1502.01589 [astro-ph.CO]).
	
	\bibitem{T. Padmanabhan2003} T. Padmanabhan, Cosmological Constant - the Weight of the Vacuum, {\it Phys. Repts.} \textbf{380} (2003), 235 (arXiv:hep-th/0212290).
	
	\bibitem{S. Weinberg1989} S. Weinberg, The cosmological constant problem, {\it Rev. Mod. Phys.} \textbf{61} (1989), 1.
	
	\bibitem{V. Sahni2000} V. Sahni, and A. A. Starobinsky, The Case for a Positive Cosmological Lambda-term, {\it Int. J. of Mod. Phys.} \textbf{9}(2000),373 (arXiv:astro-ph/9904398).
	
	\bibitem{I. Zlatev1999} I. Zlatev, L.-M. Wang, and P. J. Steinhardt, Quintessence, Cosmic Coincidence, and the Cosmological Constant, {\it Phys. Rev. Lett.} \textbf{82}(1999), 896 (arXiv:astro-ph/9807002).
	\bibitem{Copeland} E. J. Copeland, M. Sami and S. Tsujikawa, Dynamics of Dark Energy, Int. J. Mod. Phys. D \textbf{15} (2006), no.11, 1753-1935 (arXiv:hep-th/0603057).	
	
	\bibitem{Sujay2017} S. Kr. Biswas, W. Khyllep, J. Dutta and S. Chakraborty, Dynamical analysis of an interacting dark energy model in the framework of a particle creation mechanism, {\it Phys. Rev. D} \textbf{95} (2017), 103009  (arXiv:1604.07636 [gr-qc]).
	
	\bibitem{Yuri.L.Bolotin2014} Y. L. Bolotin, A. Kostenko, O. A. Lemets and D  A. Yerokhin, Cosmological Evolution With Interaction Between Dark Energy And Dark Matter,  {\it Int. J. Mod. Phys. D} \textbf{24} (2015), no.03 1530007 (arXiv: 1310.0085 [astro-ph.CO]).
	
	\bibitem{M.Khurshudyan2015}  M. Khurshudyan and R. Myrzakulov, Phase space analysis of some interacting Chaplygin gas models, {\it Eur. Phys. J. C}  \textbf{77} (2017), 65 (arXiv:1509.02263 [gr-qc]).
	
	\bibitem{S.Kr.Biswas2015a} 	S. Kr. Biswas and S. Chakraborty, Dynamical systems analysis of an interacting dark energy model in the brane scenario, {\it Gen. Rel. Grav.} \textbf{47} (2015), 22 (arXiv:1502.06913 [gr-qc]).
	
	\bibitem{S.Kr.Biswas2015b}  S. Kr. Biswas and S. Chakraborty, Interacting Dark Energy in f(T) cosmology : A Dynamical System analysis, {\it Int. J.  Mod. Phys. D.} \textbf{24} (2015), no.07, 1550046.
	
	\bibitem{N.Tamanini2015} N. Tamanini, Phenomenological models of dark energy interacting with dark matter {\it Phys. Rev. D} \textbf{92} (2015), no.4, 043524 (arXiv: 1504.07397 [gr-qc]).
	
	\bibitem{Xi-ming Chen2009} X. Chen, Y. Gong and E.N. Saridakis, Phase-space analysis of interacting phantom cosmology, {\it JCAP} \textbf{0904} (2009), 001 (arXiv:0812.1117[gr-qc]).
	
	\bibitem{T.Harko2013} T. Harko and F. S. N. Lobo, Irreversible thermodynamic description of interacting dark energy-dark matter cosmological models, {\it Phys. Rev. D} \textbf{87} (2013), no.4, 044018 (arXiv:1210.3617[gr-qc]).
	
	
	\bibitem{Wang2016} B. Wang, E. Abdalla, F. Atrio-Barandela and D. Pavon, Dark Matter and Dark Energy Interactions: Theoretical Challenges, Cosmological Implications and Observational Signatures, {\it Rept. Prog. Phys.} {\textbf 79} (2016), no.9, 096901 (arXiv:1603.08299 [astro-ph.CO]).
	
	
	\bibitem{Copeland1} E. J. Copeland, A. R. Liddle and D. Wands, Exponential potentials and cosmological scaling solutions, {\it Phys. Rev. D} \textbf{57} (1998), 4686-4690 (arXiv:gr-qc/9711068).
	
	
	\bibitem{Fang1} W. Fang, Y. Li, K. Zhang, and H. Qing Lu,  Exact Analysis of Scaling and Dominant Attractors Beyond the Exponential Potential, {\it Class. Quant. Grav.} \textbf{26 } (2009), 155005 (arXiv:0810.4193 [hep-th]).
	
	\bibitem{Leon1} G. Leon, On the Past Asymptotic Dynamics of Non-minimally Coupled Dark Energy, {\it Class. Quant. Grav.} \textbf{26} (2009), 035008 (arXiv:0812.1013 [gr-qc]).
	
	\bibitem{Honorez} L.L. Honorez, O,Mena	and G. Panotopoulos, Higher order Coupled Quintessence, {\it Phys.Rev.D} \textbf{82} (2010),123525 (arXiv: 1009.5263 [astro-ph.CO])
	
	\bibitem{Odintsov2018b} S. D. Odintsov and V. K. Oikonomou, Study of finite-time singularities of loop quantum cosmology interacting multifluids, {\it Phys. Rev. D} \textbf{97} (2018), 124042 (arXiv:1806.01588 [gr-qc]).
	
	\bibitem{Aljaf} M. Aljaf, D. Gregoris, M. Khurshudyan, Phase space analysis and singularity classification for linearly interacting dark energy models, {\it Eur. Phys. J. C} \textbf{80} (2020), 112.
	
	\bibitem{S. Bahamonde} S. Bahamonde, Generalised nonminimally gravity-matter coupled theory, {\it Eur. Phys. C} \textbf{78} (2018), 326 (arXiv:1709.05319[gr-qc]).
	
	\bibitem{V.K. Oikonomou} V.K. Oikonomou, Generalized Logarithmic Equation of State in Classical and Loop Quantum Cosmology Dark Energy-Dark Matter Coupled Systems, { \it Annals Phys} \textbf{409} (2019), 167934.
	
	\bibitem{Biswas2021} S. Kr. Biswas and A. Biswas, Phase Space Analysis and Thermodynamics of Interacting Umami Chaplygin Gas in FRW Universe, {\it Eur. Phys. J. C} \textbf{81} (2021), 356. (arXiv:2011.05760[physics.gen-ph])
	
	\bibitem{Mandal2022} G. Mandal, S. Kr. Biswas, S. Saha and A. Al Mamon, Dynamical system analysis of logotropic dark fluid with a power law in the rest-mass energy density, {\it Phys. of the Dark. Univ.} \textbf{35} (2022), 100970 (arXiv: 2107.01050[gr-qc]).
	
	
	
	\bibitem{Bahamonde2018} S. Bahamonde, C. G. Boehmer, S. Carloni, E. J. Copeland, W. Fang and N. Tamanini, Dynamical systems applied to cosmology: Dark energy and modified gravity, {\it Phys. Rept.} \textbf{775-777} (2018), 1-122  (arXiv:1712.03107 [gr-qc]).
	
	\bibitem{Anderson} G. W. Anderson and S. M. Carrol, Dark matter with time dependent mass, based on a talk by SMC at Cosmo-97  (World Scientific), (1998) pp. 227-229, 
	(arXiv:astro-ph/9711288).
	
	
	\bibitem{Damour} T. Damour, G.W. Gibbons, C. Gundlach, Dark Matter, Time Varying GG, and a Dilaton Field, {\it Phys. Rev. Lett.} 64 (1990), 123.
	
	
	\bibitem{Zhang} X. Zhang, Coupled Quintessence in a power-Law case and the cosmic coincidence Problem, {\it Mod.Phys.Lett. A} \textbf{20} (2005), 2575 (arxiv:astro-ph/0503072).
	
	\bibitem{Comelli} D. Comelli, M. Pietroni and A. Riotto, Dark Energy and Dark Matter,{\it Phys.Lett.B} \textbf{571} (2003), 115-120 (arXiv:hep-ph/0302080).
	
	\bibitem{Franca} U. Franca and R. Rosenfeld, Age constraints and fine tuning in variable-mass particle models, {\it Phys.Rev.D} \textbf{69} (2004), 063517 (arXiv: astro-ph/0308149).
	
	\bibitem{G.Leon-Saridakis} G. Leon and E. N. Saridakis, Phantom dark energy with varying mass dark matter particles: Acceleration and coincidence problem, {\it Phys. Lett. B} \textbf{693} (2010), 1-10.
	
	\bibitem{Soumya} S. Chakraborty, S. Mishra and S. Chakraborty, A Dynamical System Analysis of cosmic evolution with coupled phantom dark energy with dark matter, {\it International Journal of Modern Physics D}, \textbf{31} (2022), No. 01, 2150129 (arXiv: 2011.09842[gr-qc]).
	
	\bibitem{Felice} A. De Felice and S. Tsujikawa, f(R) theories, {\it Living Rev. Rel}, \textbf{13} (2010), 3 (arXiv:1002.4928 [gr-qc]).
	
	\bibitem{Nojiri} S. Nojiri and S. D. Odintsov, Modified Gauss-Bonnet theory as gravitational alternative for dark energy, {\it Phys. Lett. B} \textbf{631} (2005), 1 (	arXiv:hep-th/0508049).
	
	\bibitem{Weyl} H. Weyl, {\it Sber. preuss. Akad. Wiss. (Berlin)}, \textbf{1918}  (1918), 465. 
	
	\bibitem{Lyra} G. Lyra, {\it Math. Z.}, \textbf{54} (1951), 52.
	
	\bibitem{Sen1957} D. K. Sen, A Static Cosmological Model, {\it Z. Phys.} \textbf{149} (1957), 311.
	\bibitem{Halford} W. D. Halford, Cosmological theory based on Lyra's geometry, {\it Aust. J. Phys.} \textbf{23} (1970), 863-9.
	
	\bibitem{Beesham} A. Beesham,  FLRW cosmological models in Lyra's manifold with time dependent displacement field, {\it Aust. J. Phys.} \textbf{41} (1988), 833-42.
	
	\bibitem{Darabi} F. Darabi, Y. Heydarzade and F. Hajkarim, Stability of Einstein Static Universe over Lyra Geometry,{\it Can. J. Phys.} \textbf{93} (2015), 1566 (arXiv:1406.7636 [gr-qc]).
	
	\bibitem{Kangujam P Singh} K.P. Sing, M.R. Mollah, Could the Lyra manifold be the hidden source of the dark energy?, {\it Int. J. Geom. Methods Mod. Phys.} {\textbf Vol. 14}(2017), No. 04, 1750063 (arXiv:1901.11394  [physics.gen-ph]). 
	
	\bibitem{Hoavo} H. Hova, A dark energy model in Lyra manifold, {\it J. Geom. Phys.} \textbf{64} (2013), 146 (arXiv:1204.0774 [gr-qc]).
	
	\bibitem{Shchigolev} V.K. Shchigolev, E.A. Semenova, Scalar Field Cosmology in Lyra's Geometry, {\it International Journal of Advanced Astronomy}, {\textbf 3(2)}(2015), 117-122, (arXiv:1203.0917 [gr-qc]).
	
	\bibitem{Khurshudyan-Lyra} M. Khurshudyan, J. Sadeghi, R. Myrzakulov A. Pasqua and H. Farahani, Interacting Quintessence Dark Energy Models in Lyra Manifold, { \it Advances in High Energy Phys.} \textbf{2014} (2014), 878092.
	
	\bibitem{Cuzinatto} R. R. Cuzinatto, E. M. de Morais and B. M. Pimentel, a Scalar-Tensor Theory of Gravity on Lyra Manifold, { \it Phys. Rev. D} \textbf{103} (2021), 124002 (arXiv:2104.06295 [gr-qc]).
	
	\bibitem{Sen1971} D. K. Sen and K. Dunn, A Scalar-Tensor Theory of Gravitation in a Modified Riemannian Manifold, {\it J. Math. Phys. (N.Y.)} \textbf{12} (1971), 578.
	
	\bibitem{Bhamra} K. Bhamra, A Cosmological Model of Class One in Lyra's Manifold, {\it Aust. J. Phys.} \textbf{27} (1974), 541.
	\bibitem{Beesham2} A. Beesham, Friedmann's cosmology in Lyra's manifold, {\it Astrophys. Space Sci.} \textbf{127} (1986), 355.
	\bibitem{Reddy} D. R. Reddy and R. Venkateswarlu, { \it Astrophys. Space Sci.} \textbf{149} (1988), 287.
	\bibitem{Singh1} G. P. Singh and K. Desikan, A new class of cosmological models in Lyra's geometry, {\it Pramana}  \textbf{49} (1997), 205.
	
	\bibitem{Singh2} J. K. Singh, Exact solutions of some cosmological models in Lyra geometry, {\it Astrophys. Space Sci.} \textbf{314} (2008), 361.
	
	\bibitem{Saadat} H. Saadat, A Cosmological Model of the Early Universe Based on ECG with Variable $\Lambda-Term$ in Lyra Geometry, {\it Int. J. Theo. Phys.} \textbf{55} (2016), 2364 [arXiv:1508.06544].
	
	
	
	
	
	\bibitem{Coley} A. A. Coley, Dynamical systems and cosmology. {\it (Kluwer Academic Publishers, Dordrecht Boston London, 2003)}. 
	

	
	\bibitem{Piazza} F.Piazza, S.Tsujikawa, Dilatonic ghost condensate as dark energy, {\it JCAP} \textbf{0407} (2004), 004  (arXiv:hep-th/0405054).
	
	\bibitem{Mahata} N. Mahata, S. Chakraborty, Dynamical System Analysis for a phantom model, {\it Gen. Relt. Grav.} \textbf{46} (2014), 1721 (arXiv:1312.7644 ).
	
	\bibitem{Garriga} J. Garriga and V. F. Mukhanov, Perturbations in k-inflation, {\it Phys. Lett. B.} \textbf{458} (1999), 219.
	
	\bibitem{Sudipta Das} S. Das, A. Al MAmon and M. Banerjee, A new parametrization of dark energy equation of state leading to double exponential potential, Research in Astronomy and Astrophysics \textbf{18} (2018), 131 (arXiv:1805.07148 [gr-qc]).
	


\end{thebibliography}
\end{document}